\documentclass[10pt,letterpaper]{article}

\usepackage{jheppub}
\usepackage[T1]{fontenc} 
\usepackage{here}
\usepackage{grffile}
\usepackage{subfigure}
\usepackage{color}

\title{DO WE FIND HIGH ENERGY PHYSICS INSIDE (ALMOST) EVERY 
SOLID OR FLUID AT LOW TEMPERATURE?}
\author[a]{H. B. Nielsen}
\author[b]{and M. Ninomiya}

\affiliation[a]{Niels Bohr Institute, University of Copenhagen,\\17 Belgdamsvej, DK 2100 Copenhagen $\phi$, Denmark}
\affiliation[b]{Advanced Mathematical Institute Osaka-city University, Sugimoto 3-3-138, Sumiyoshi-ku Osaka, 1558-8585, Japan\\  and Yukawa Institute for Theoretical Physics, Kyoto University Kyoto. 606-8502, Japan }

\abstract{It is an old idea of ours (H. B. Nielsen ``Dual Models'' section 6 ``Catastrophe 
Theory Program'' Scottish University Summer 
School 
1976) that a most general material with only 
translation 
symmetry, but otherwise no symmetries should 
generically 
(in general) have some small regions in quasi momentum space, 
where you ``see'' an approximate Weyl equation behavior. The 
Weyl equation is the relativistic equation for a (left handed) 
neutrino. This remark means that one could imagine, that there 
were behind the Standard Model of High energy physics, a very 
general crystal model with very little symmetry. Even for the 
Yang Mills or electrodynamics types fields a similar 
philosophy is possible. There are though some problems with 
this solid-state type of model beyond the Standard model, for 
which we thought have some remedy by means of homolumo gap 
effects. 

By making use of relativistic quantum field theory on the 
lattice
we predicted 
theoretically very high magneto-conduction due to 
Adler-Bell-Jackiw chiral anomaly effect -- so called 
Nielsen-Ninomiya effect (or mechanism) in gapless parity 
violating material. Nowadays this kind of material such as 
chiral or Weyl semimetal and the effect are detected by 
experiments.  

\hrulefill

}
\begin{document}
\maketitle
\section*{Introduction}


The authors, in particular H. B. N. have through many years 
the dream, that it is {not 
important   
  what the (most) fundamental 
laws of Nature might be}, because almost certainly the 
same effective laws would come out anyway: This philosophy is 
called ``Random 
Dynamics''.

Inside a piece of matter - crystal, glass, ... -
one should then at {very low temperature } 
according to this dream find the Standard Model.

{Recently one is about to find Cases of 
Relativity-behaving Quasi-particles:}
A material, e.g. graphene, with such simulations of 
relativistic 
particles as we talk about. 

Materials with relativistic 
particles simulated as quasiparticles may be very 
applicable to say high conductivity purposes,... 

{Some of our publications:}
\begin{itemize}
\item
  H.~B.~Nielsen and M.~Ninomiya,
  ``No Go Theorem for Regularizing Chiral Fermions,''
  Phys.\ Lett.\  {\bf 105B}, 219 (1981).
\item
H.~B.~Nielsen and M.~Ninomiya,
  ``Absence of Neutrinos on a Lattice, 1. Proof by homotopy theory''
  Nucl.\ Phys.\ B {\bf 185}, 20 (1981).
\item
  H.~B.~Nielsen and M.~Ninomiya,
  ``Absence of Neutrinos on a Lattice. 2. Intuitive Topological Proof,''
  Nucl.\ Phys.\ B {\bf 193}, 173 (1981).
\item 
As for the initiation of Random Dynamics, See
``Fundamentals of Quark Models''. Proceedings: 17th Scottish 
Universities Summer School in Physics, St. Andrews, Aug 1976, 
I.M. Barbour, A.T. Davies (Glasgow U.);1977 - 588 pages; 
Edinburgh: SUSSP Publ. (1977);Conference: 
C76-08-01; Contributions: Dual Strings,
Holger Bech Nielsen (Bohr Inst.). Aug 1974, 71 pp.;NBI-HE-74-15
In the last section the idea of ``Random Dynamics ''
is introduced based on finding Weyl equation in ``whatever''.
     \end{itemize}
{\bf The present paper consists as part I and part II.

The part I:
Relativity Theory found in solid state.

and

The part II
``What comes beyond Topological Insulator -- 
Nielsen-Ninomiya Effect (or Mechanism) due to ABJ Anomaly --''

Part I: Relativity-Theory found in Solid State Physics} 
\begin{description}
\item[I-1]{\color{blue}Introduction} 
\item[I-2]{\color{blue}Automatic:} a pet-thought: 
Natural laws come by themselves! (``Random Dynamics'')
\item[I-3]{\color{blue} General:} 
A very general world with (only) momentum conservation.
\item[I-4]{\color{blue} Graphene:} Example Graphene.
\item[I-5]{\color{blue} Heusler:} Half-metals, Heusler compounds.
\item[I-6]{\color{blue} Wang:} 
Thoughts about making materials having models of relativistic
particles inside.
\item[I-7]{\color{blue} Doubling:} Nielsen - Ninomiya theorem 
about 
doubling of such relativistic particles unavoidably on the 
lattice.
great future;
hope of seeing high energy physics in low temperature  
materials not out, but not quite finished.
material simulates relativistic quantum field theory.    
\item[I-8]{\color{blue} Further}: Further Developments of our 
``Random Dynamics''
\item[I-9]{\color{blue} Conclusion for part I}
\end{description}

The part II: What comes beyond Topological Insulator 
--Nielsen-Ninomiya Effect (or Mechanism) due to ABJ Anomaly

\begin{description}

\item[II-1]: {\color{blue} Introduction}
\item[II-2]: {\color{blue} 1+1 dimensional Example}
\item[II-3]: {\color{blue} 3+1 dimensional case} Weyl 
(or chiral) Fermion Adler-Bell-Jackiw Anomaly
\item[II-4]: {\color{blue} Parity non-invariant, Zero-gap 
material}
\item[II-5]: {\color{blue} Transfer from Left- to Right- comes 
by Adler-Bell-Jackiw anomaly}
\item[II-6]: {\color{blue} Further arguments}
\item[II-7]: {\color{blue} Conclusions}
\item[Appendix A]: {\color{blue} Necessary properties of 
quantum field theory in this paper}
\item[Appendix B]: {\color{blue} Adler-Bell-Jackiw anomaly in 
continuum spacetime}

\end{description}

\section*{I-2  Automatic}

{\bf Our Old Work in 1976: 
Dreams Laws of Nature Automatic}

{\color{green}"Dual Strings. Fundamentals of Quark Models."
by H. B. Nielsen, in Scottish University Summer School in 
Physics, 
St. Andrews, 1976} (There H.B.N. still mainly is talked on 
String theory,
but at the end a general (fermion) Hamiltonian is studied.)

Assumed was {\bf \color{blue} translational invariance}, at 
least 
with 
respect to a lattice say, and thus a (quasi) momentum 
conservation, but with respect to the {\bf \color{blue}
``internal degrees 
of freedom''} there is a {\bf \color{blue}very general} 
theory, though 
assuming there being essentially a finite (discrete). system 
of states(representing possibly spin and band degrees 
of freedom.).

{\bf (Trivial) Generic Considerations on Fermion 
Dispersion relations (1976)}. We ignore all conservation 
laws except for
\begin{itemize}
\item Energy conservation and Hamiltonian development.
\item Momentum Conservation.
\item Particle (number) conservation.
\item Free approximation (first).
\item Smoothness, (so that e.g ${\bf H}(\vec{p})$ is 
differentiable and continuous as function of $\vec{p}$.)
\item Generic: i.e. no fine-tuned values of parameters,   
\end{itemize} 
and consider a single particle equation:
\begin{equation}
i\frac{\partial}{\partial t}\psi(\vec{p},t) ={\bf H}(\vec{p})
\psi(\vec{p}),
\end{equation}  
where for each value of the momentum $\vec{p}$ the ${\bf H}
(\vec{p})$ is a Hermitian {matrix}.

{Relativity and Dimensionality of Space time being 3+1 come 
out Automatically!}

A priori - with no fine-tuning (=generically) - the Fermi 
surface
would put itself at separate eigenvalues; but if for 
some reason ( e.g. ``homlumo-gap effect'') the Fermi-level
were {just where $n=2$ levels meet}, then in a small
neighborhood the shape of the dispersion relations would 
be given by taking ${\bf H(\vec{p})}$ to be 
$n\times n = 2\times 2$. We then Taylor expand
\begin{eqnarray}
{\bf H}(\vec{p}) &\approx & {\bf H}(\vec{p}_0) + 
\sum_{a,\mu} \sigma^a  V^{\mu}_a p_{\mu}+...  
\end{eqnarray} 
where $\sigma^a$ are the Pauli-matrices and the unit 
matrix $\sigma^0 = {\bf 1}$. The ``vierbein'' $V^{\mu}_a$
is a set of expansion coefficients for ${\bf H}(\vec{p})$
as function of the components $p_{\mu}$ (strictly speaking 
$\mu$ =1,2,3; here).

{\bf Hermitian matrix, Provided Fermi-level at Degeneracy $n=2$
leads to Weyl Equation in 3+1 Dimensions.}

In the old days we argued that in a general physics universe 
the {\bf \color{red}Hubble expansion} would finally lead to 
the Fermi-level 
approaching an $n=2$ degenerate levels energy; but now
H. B. N.'s Zagreb group - I.Andric, L. Jonke, D. Jurman, 
and HBN - have studied 
in general, what is called {\bf\color{red} 
``Homolumo-gap Effect''} meaning 
the by Jahn and Teller\cite{JT} first proposed effect, that the 
electrons filling the Fermi-sea would back react such as 
to increase the homolumo gap between the lowest unoccupied
(LUMO) and the highest occupied (HOMO) state. This effect 
goes in the direction to make metals not occur, and 
make every materials become an insulator, but the 
gapless semiconductor may be too hard for the homolumo-gap 
effect to dispense with.    

Note that this hope for getting automaticly a Weyl-equation
like theory had, when using just  Hermitean Hamiltonian 
marices and looking at the $n=2$ degeneracy possibility, the
consequence that there came only {\em three} spatial 
dimensions functioning the relativistic way, because 
there were only 3 Pauli matrices. Somehow arguing that 
the dimensions for which there are no Pauli matrices 
will lead to essentially zero velocity for the 
fermion/quasi-electron in these directions and that 
such dimensions will not be observed, we have come to 
3+1 dimensions as an additional prediction from 
the very general starting theory!   

With time-reversal symmetry imposed dimension 
prediction gets modified.

\begin{table}[h]
\begin{tabular}{|c|c|c|c|c|}
\hline
Symmetry&Square&Pauli M.& Dimension&Field\\
\hline
TP& $(TP)^2=1$& $\sigma_x, \sigma_z$& 2+1& Real\\
-&-& $\sigma_x,\sigma_y,\sigma_z$& 3+1& Complex\\
TP & $(TP)^2 = -1$& 5 of them&5+1& Quaternions\\
\hline
\end{tabular}
\caption{The symmetry assumed in line 1 and 3 is the 
combination of {time reversal} T and parity P to 
TP, which leaves the momentum $\vec{p}$ invariant but 
is an {antilinear} operator effectively conjugating 
the complex numbers in the matrix. If then Fermi-level 
falls at $n=2$ degenerate levels 
in addition to the Kramers-Kronig doubling in the 3rd
case, one gets by Taylor expanding 
the $2\times 2$ resolved into Pauli-matrices, and a generalized
Weyl equation results corresponding to the in fourth column 
denote {\em space + time} dimensions. Actually the effective 
theory is naturally written in terms of the in column 5 
mentioned division-algebra(= field).}  
\end{table} 

{\bf  Fundamentally in many Dimensions, but in 
Most dimensions the Fermion Run with Zero Velocity, we 
Ignore them.}

\vspace{3mm}
In the for fundamental physics ideal situation of 
{\bf\color{red} no extra $T$ or $TP$ symmetry } the Hamiltonian 
matrix ${\bf H}(\vec{p})$ is just a generic($\sim$ random)
Hermitian matrix (with {complex} matrix elements),
and it predicts at the two levels degenerate point 
- hoped to be favored at the Fermi-surface by 
either Hubble expansion or homolumo-gap-effect - that 
the Fermion only moves with appreciable velocity in 
as many spatial dimensions as there are Pauli-matrices.
We hope that the dimensions in which the velocity gets zero,
can/shall be ignored. If the zero-velocity dimensions 
are ignored, then we have remarkable agreement:

{The number of dimensions in which the generic 
double degeneracy neighborhood has the fermions move 
just corresponds to experimental number of dimensions
3+1 and to having relativity and rotational invariance!} 


{\color{blue}{\bf If } TP (or T) {\bf is good symmetry and} $(TS)^2 = 1${\bf then}  ${\bf H}(\vec{p})$ {\bf 
must have real matrix elements.}} 

This is the case in which we in a crystal - with 
PT symmetry say - {\bf completely ignore} the usual
{spin} as being decoupled so as to be totally 
ignored. 

In this case we get the effective dimensionality, if we ignore 
the zero-velocity directions:

\begin{center}
{\bf\color{blue} \huge 2 + 1 }
\end{center}

This means that the relativistic effective fermion 
should appear ``generically'' (automatically) even in 
only 2 spatial dimensions.

{With Genuine Spin=$\frac{1}{2}$ Electrons and Unbroken
Time reversal, the ``Quaternion Case''}
{If} $T$ or $TP$ {good symmetries, and spin 
$\frac{1}{2}$ included, then} $T^2 = (TP)^2 =-1$    
{we have generally doubling of all levels according 
to Kramers-Kronig rule.\cite{KK}}

So double degeneracy is already there generally and nothing
special. In this case we shall therefore instead 
consider that we can get 4 times degenerate levels 
sporadically. If we go to such a 4-times degenerate 
point in momentum space, we could elegantly go to 
a quaternion $2\times 2$ matrices (quaternions are 
writable as $2\times 2$ complex matrices, so that 
$2 \times 2$ quaternion matrices can be equivalent 
to $4 \times 4$ complex matrices with some restriction.
Dimension of non-zero velocity directions:
\begin{center}
{\bf \huge \color{blue} 5 + 1}
\end{center}

\section*{I-3  Graphene}
Graphene denotes the layer of carbon like the ones 
in graphite taken as seperate, i.e. it is 2(space)dimensional
material. The quasi electrons running in the graphene layers
actuall do show dispersion relations behaving how we 
above argued for the case with time reversal but 
ignoring the spin leading to the effective 
space time dimension 2+1. 

On the following picture one sees the lattice structure 
of graphene:


\begin{center}
\includegraphics{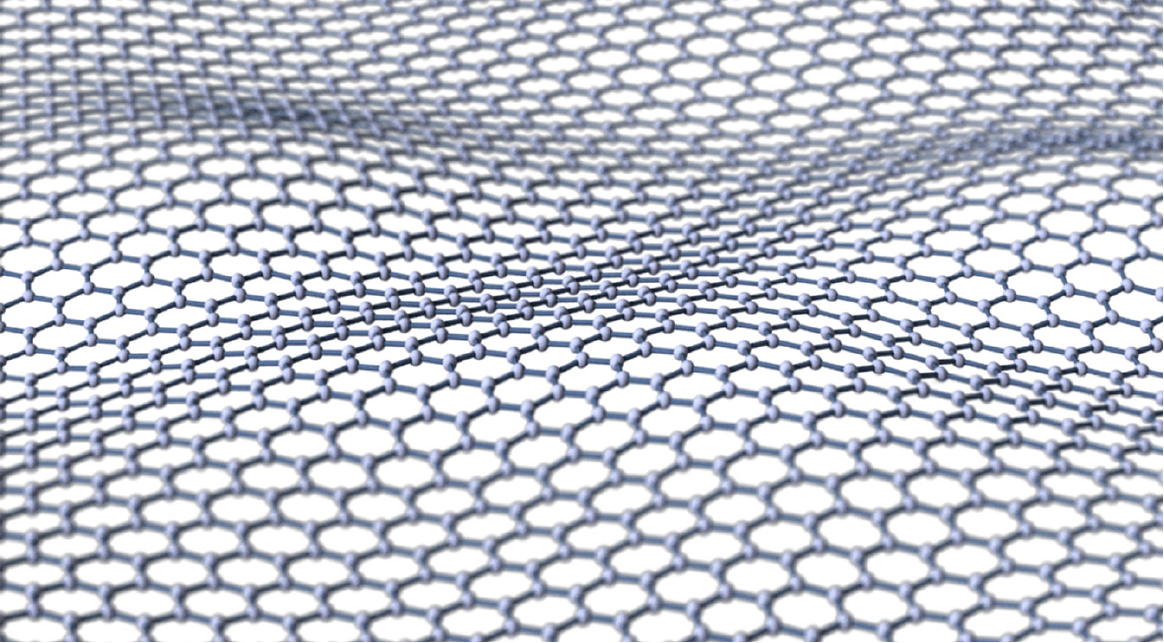}

{\bf (2+1)-dimensional Example is Graphen.}
\end{center}
The next figure is supposed to generally illustrate 
a metal, an insulator and a material with a Dirac-like
quasi particle (on the figure graphen).

\begin{center}
\includegraphics{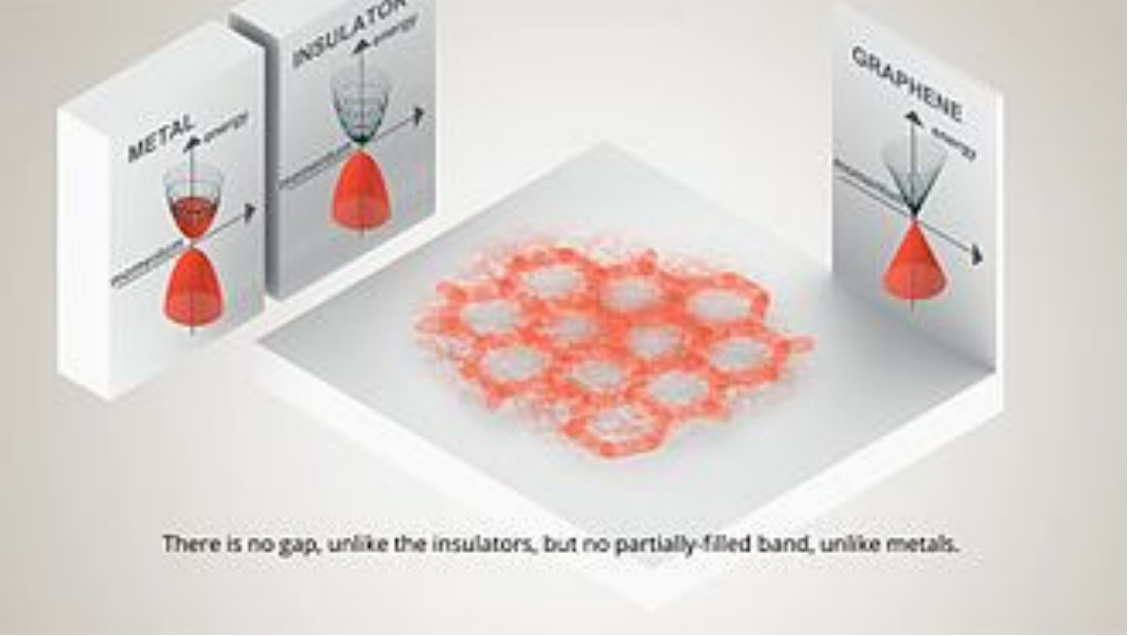}

Even just making a two-layer of graphene complicates 
the situation and the work by Gammelgaard on the next 
figure illustrates a gap appearing:

{\bf Putting Double Layer Produces Gap}
\includegraphics{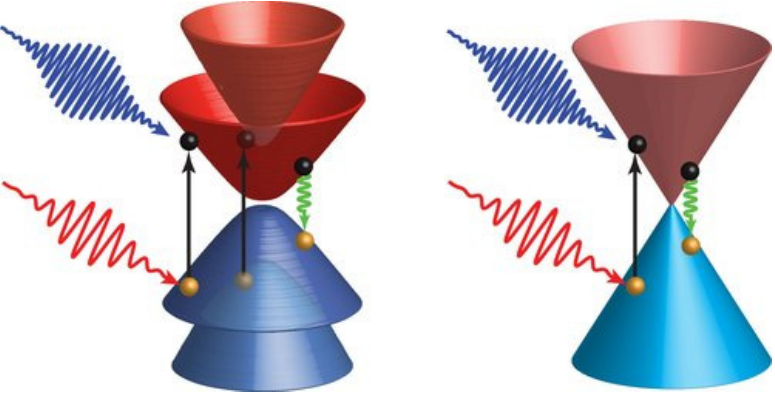}

The left dispersion law is for a double layer of 
graphene; the right for single layer. (Gammelgaard).
\end{center}

The next figures illustrate calculation of the 
dispersion relations for quasi-electrons in graphene 
by the model described just below. Since we have a 2 space 
dimension material the energy can be the orbital direction 
up in the perspective while the two spatial momentum components
form the basis plane of the three-dimensional perspective
figure:

\includegraphics[clip, width=8cm]{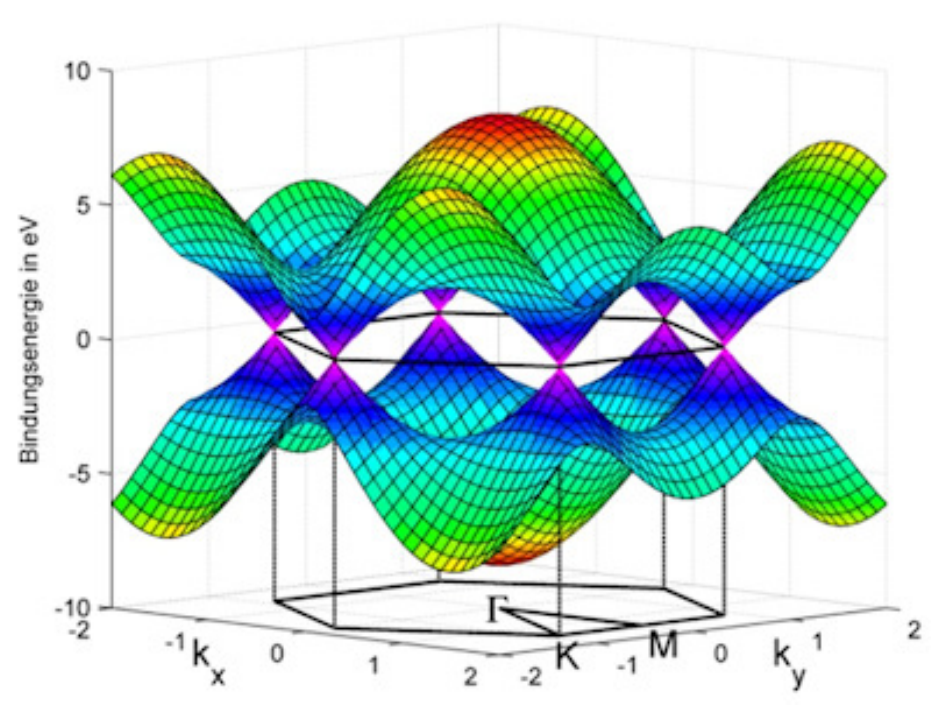}
\includegraphics[clip, width=8cm]{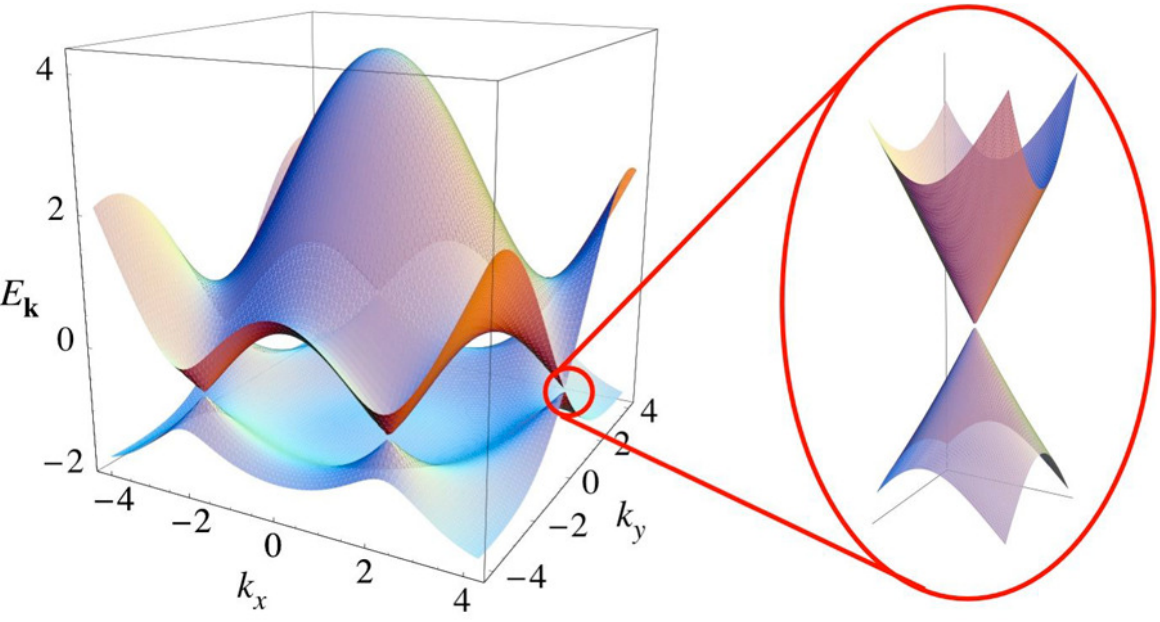}
\
The Dirac points are  of course the points where two 
branches of the dispersion relation meet {\em with 
a cone shape.}:

\begin{center}
\includegraphics{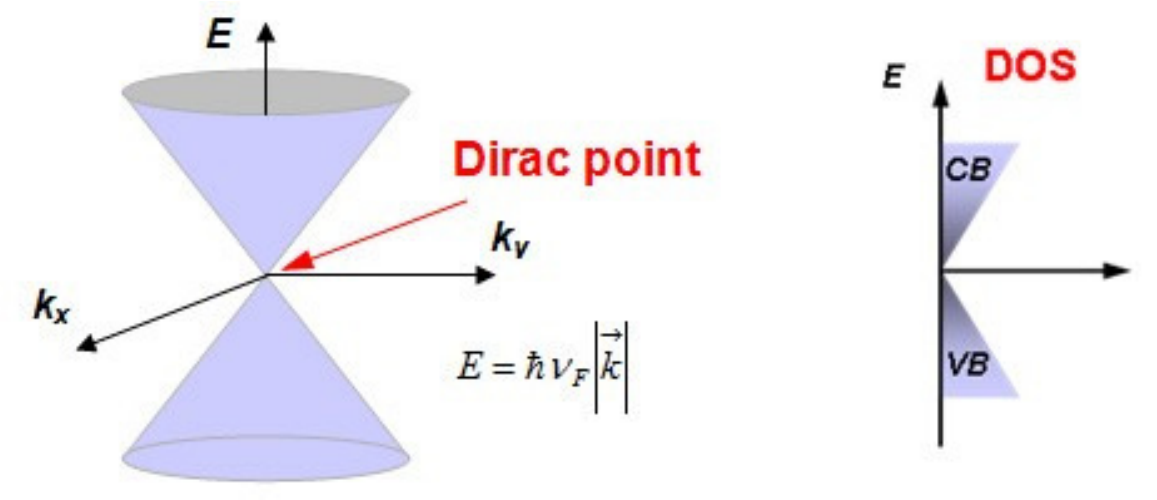}
\end{center}
%

{\bf Dispersion Relation of Graphene}
The electronic properties of graphene can be described using 
a simple tight binding  model. The electrons in the covalent 
bonds form deep fully filled valence bands, and
thus their effects on the conductivity can be safely 
disregarded. The unhybridized
p orbital is only slightly perturbed by the neighboring atoms. 
Therefore, the wave
function of an electron in the system can be written as a 
Linear Combination of
Atomic Orbitals (LCAO). Using these orbitals as the basis set 
to represent the wave
function, the Hamiltonian that governs the dynamics of the 
electron is given by:
\begin{eqnarray}
H& =& \sum_i\epsilon_i|\psi_i><\psi_i|+\sum_l \sum_{\{<i|j>\}_l}t_l
(|\psi_i><\psi_j|+|\psi_j><\psi_i|)
\end{eqnarray}
where$\epsilon_i$ represents the onsite energy at the atom, 
$|\psi_i>$ the i'th atomic orbital 
, $\{<i, j>\}_l$  the set of couples of lth-nearest 
neighbors, and $t_l$ the hopping
parameter between them. 

{\bf In Graphene the Fermi- surface just Lies at the Double 
degenerate Point}

So in graphene by symmetry one really get a simulation of 
a 2+1 dimensional massless Weyl/Dirac fermion, also w.r.t. 
the placing of the fermi surface.

If we think of just the generic case of a very general 
theory there will typically be {no reason why the 
fermi surface should be just at the Weyl point} (with the 
double degeneracy).

We have, however, speculated on two mechanisms, which 
might make the fermi-surface be driven towards the 
degeneracy point:
\begin{itemize}
\item If the world in question has a strong {Hubble 
expansion},
then filled states above the degeneracy point would be 
gradually emptied and holes below the degeneracy point 
would be also gradually be expanded away/attenuated.
\item ``Homolumo-gap-effect'' - meaning that the fermions 
act back onto the various degrees of freedom that can be 
adjusted in the lattice in which the fermions run. This back 
action will be so as to in the ground state arrange to 
lower the energies of filled fermi states. Thereby 
arise the so called {Homolumo-gap}, or rather it gets 
expanded by this back action ``homolumo-gap-effect''. 
In the case that we have degeneracy point that is somehow 
topologically stabilized, as one might say of the Weyl points 
discussed here, it may not be possible for the 
homolumo-gap-effect to really produce a gap. In stead we 
expect that it will only bring the fermi surface to 
coincide with the degeneracy point; that would namely 
lower the filled states as much as possible with the 
``topological ensurance'' of the degeneracy point.    
\end{itemize}

\section*{I-4  Heusler}

{\bf Heusler Compound $Mn_2CoAl$ is a Spin Gapless Semicondutor}:

 Siham Oardi, G.H. Fecher, C. Felser and J. K{\"u}bler
(arXiv:1210.0148v1 [cond-mat.mtrl-sci], 29 Sep. 2012.) 
investigated the {\bf \color{blue} Heusler compound $Mn_2CoAl$.}
They gave the article the name {\bf \color{red} Realization 
of spin gapless semiconductors: the Heusler compound 
$Mn_2CoAl$.}

In halfmetallic ferromagnets you have so to speak metal 
as far as the electrons with one direction of the spin 
is concerned, but insulator w.r.t. to the elctrons with 
the opposite spin direction. Now it may further happen 
that we instead of the metallic we get a gapless semiconductor,
namely if we have a degeneracy point as we discussed above.
Once there is effectively only one spin of the electron 
one escapes the time reversal symmetry. Thus in such 
halfmettals there is a better chance to find Weyl points.


The following figure illustrates dispersion relation along 
a piecewise straight curve in momentum space for the two 
different spin directions along the magnetization axis for 
the compound $Mn_2CoAl$.
The dispersion relation for the two different spin orientations
are printed respectively red and blue:

\begin{center}
\includegraphics{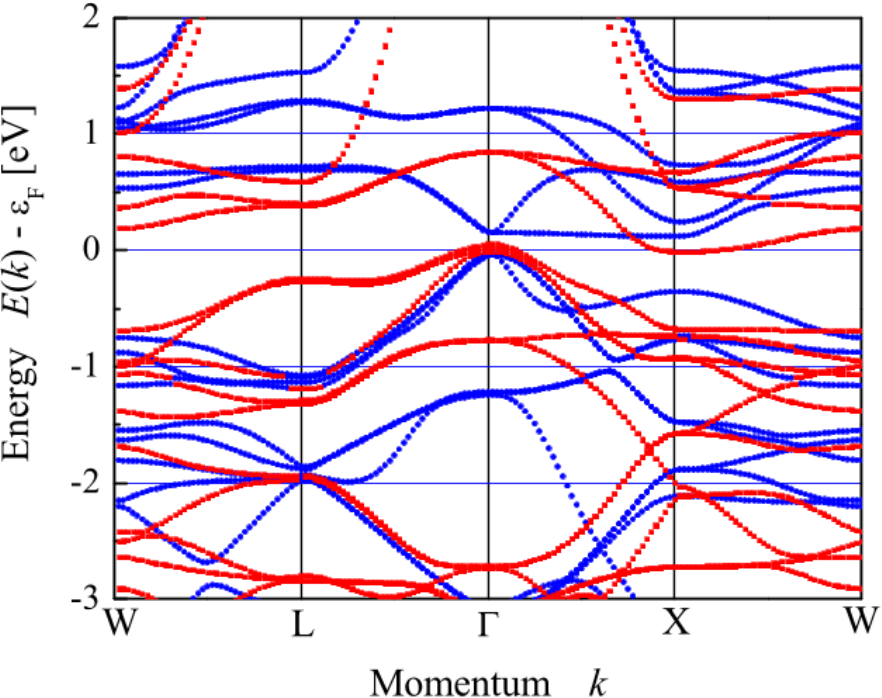}

{{\bf Band structure of $Mn_2CoAl$}, {\color{red} Majority spin 
red.}}
\end{center}


In the following three figures are then as function of 
temperature given some carrier properties of this material
$Mn_2CoAl$:

\begin{center}
\includegraphics
{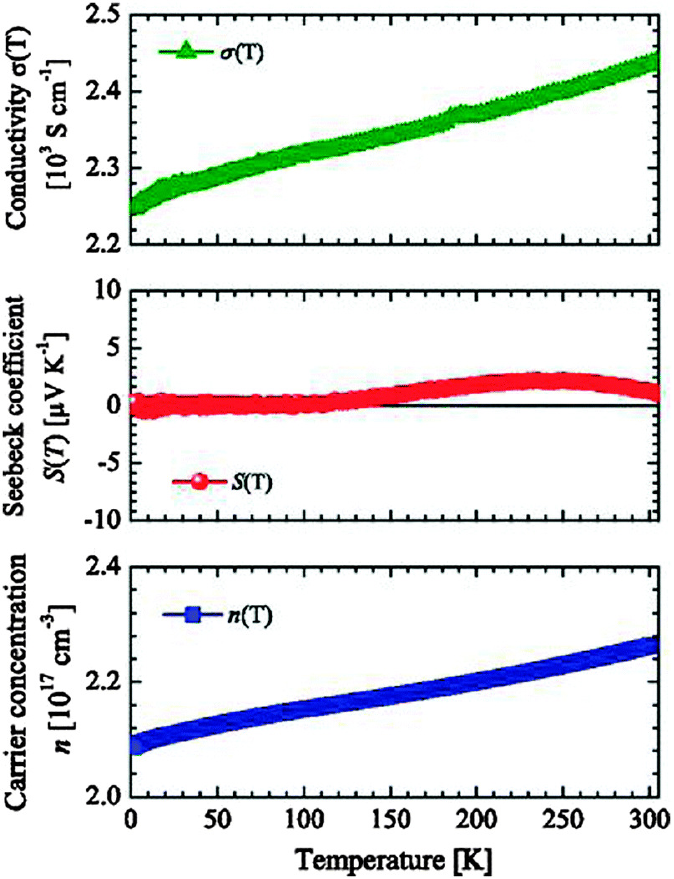}

{\color{red} Majority spin} and {\color{blue} Minority
spin}. Calculated with spin orbit coupling. 
\end{center}

On the following page from Lakhan Baisly et al. as figure 
7 in their article we see the density of electron levels 
(DOS) for the two spin orientations seperately. In the in red 
shown DOS there can be seen crudely a gap, so for this 
spin orientation we have the insulator. For the other spin 
orientation - shown with the positive ordinate pointing upwards
there is also a dip at the fermilevel, but now the DOS is going 
non-zero immediately by going away from the fermilevel. So 
for this spin we rather have the gapless semiconductor 
behavior.\\
\includegraphics{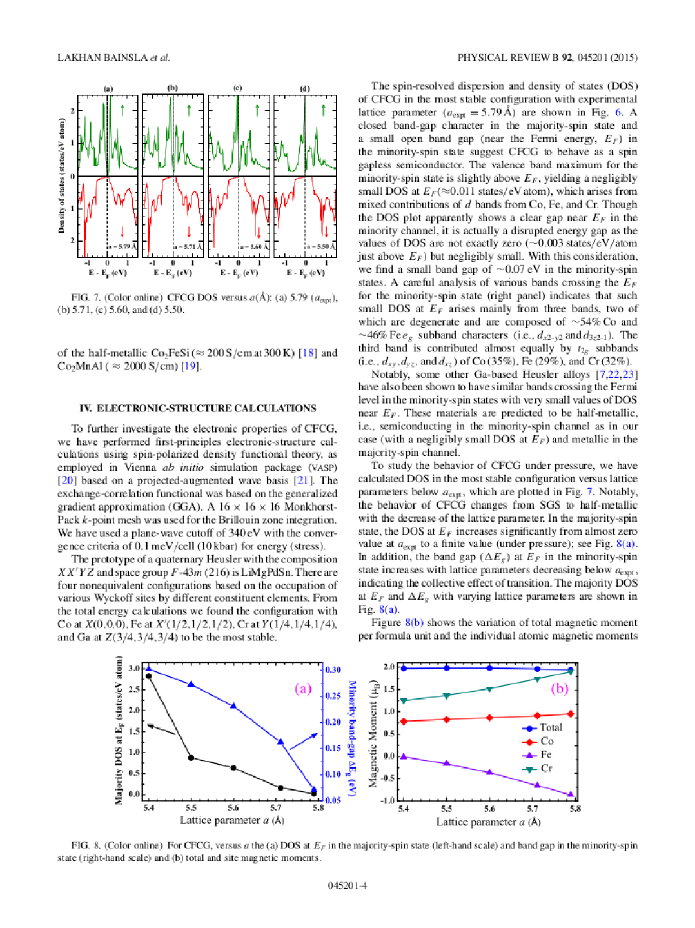}
\includegraphics{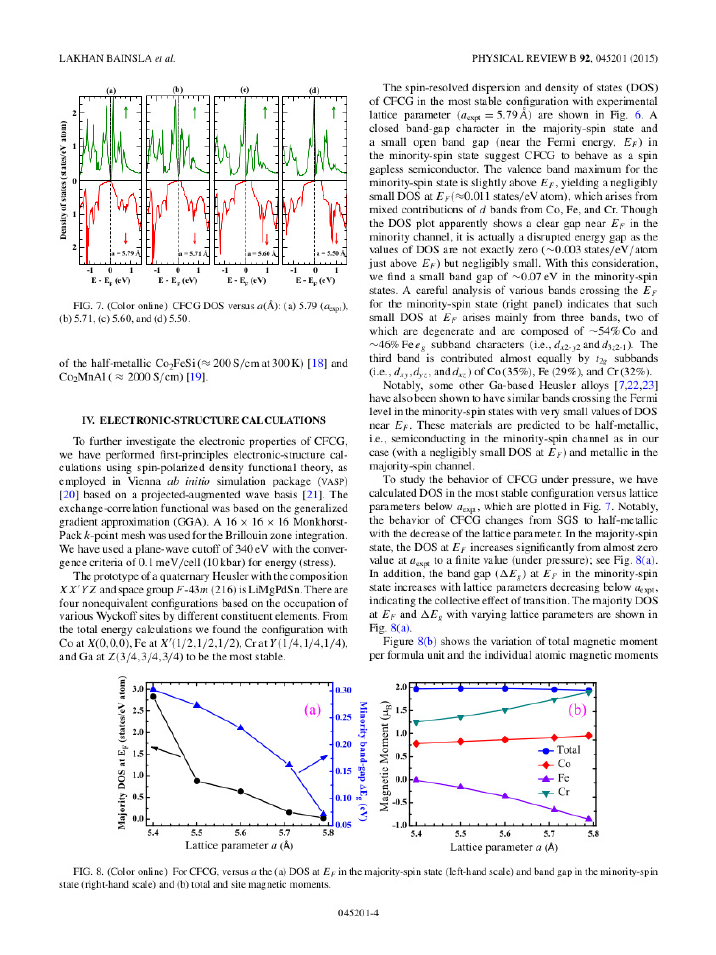}

\includegraphics{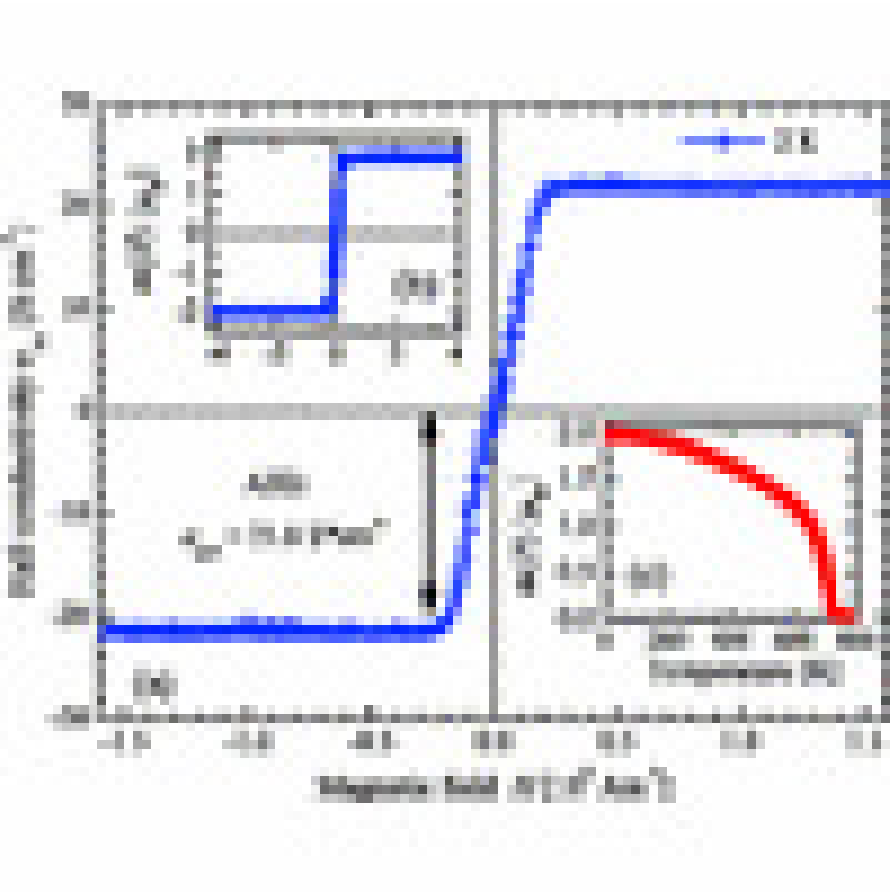}
\includegraphics{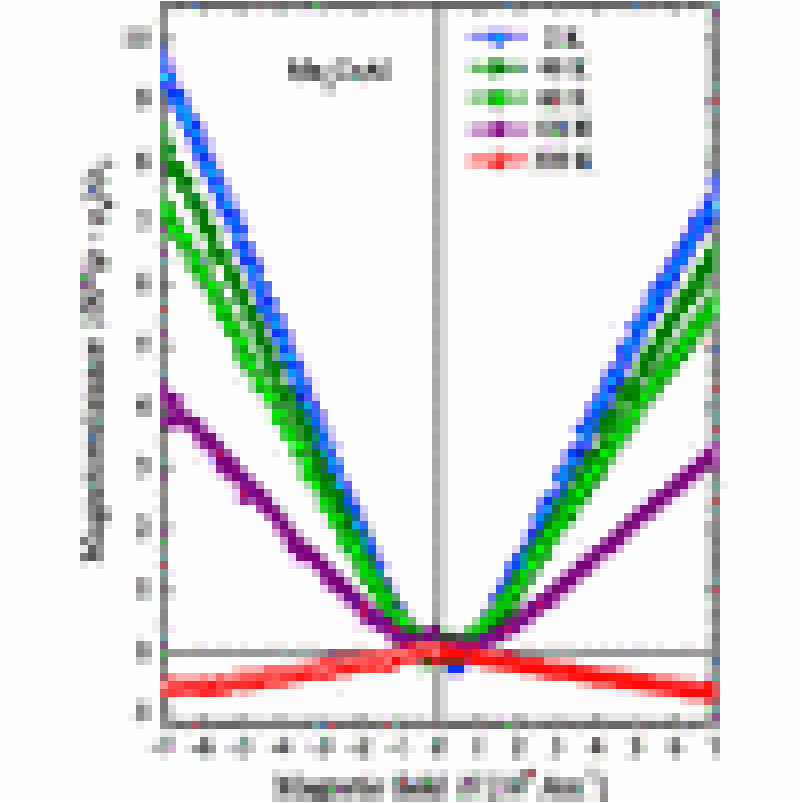}
In the figure to the left just above we see the Hall 
conductivity as function of magnetic field.

In the last just above figure we see the magnetoresistence  as function 
of a magnetic field. The strong dependence of the conductivity
as function of the magnetic field is just what one expects 
due to the Adler-Bell-Jackiw-anomaly-effect described more in 
part II of the present article below. 

These figures are from 

Siham Ouardi et al.
``Realization of Spin Gapless Semiconductors: 
The Heusler Compound ${\mathrm{Mn}}_{2}\mathrm{CoAl}$''
DOI:
10.1103/\\
PhysRevLett.110.100401

{\bf Zero Gap Material with Quadratic Energy Dispersion}
(this is by fine tuning)
$HgTe$ is one of the few materials wherin this 
{\color{red} quadratic dispersion law zero 
gap has been found, since 1950's.}

$Pb_{1-x} Sn_x Te, Pb_{1-x} Sn_x Se and Bi_x Sb_{1-x}$ are 
zero-gap materials (with quadratic disp.). 

But really one - Wang, Dou, and Zhang - expects that 
all narrow  gap 
semiconductors by  some doping or pressure  could be 
tuned to have zero gap (with quadratic dispersion law).
Then they call for finding a non-toxic material of 
this kind.

\section*{I-5  Wang}

    Physical Chemistry; Chemical Physics

Controllable electronic and magnetic properties in a 
two-dimensional germanene heterostructure
Run-wu Zhang,  Wei-xiao Ji,  Chang-wen Zhang,*  Sheng-shi Li,b  Ping Li,  Pei-ji Wang,  Feng Lia  and  Miao-juan Rena 
Author affiliations
Abstract

The control of spin without a magnetic field is one of the 
challenges in developing spintronic devices. Here, based on 
first-principles calculations, we predict a new kind of 
ferromagnetic half-metal (HM) with a Curie temperature of 
244 K in a two-dimensional (2D) germanene Van der Waals 
heterostructure (HTS). Its electronic band structures and 
magnetic properties can be tuned with respect to external strain and electric field. More interestingly, a transition from HM to 
bipolar-magnetic-semiconductor (BMS) to 
spin-gapless-semiconductor (SGS) in a HTS can be realized 
by adjusting the interlayer spacing. These findings provide 
a promising platform for 2D germanene materials, which hold 
great potential for application in nanoelectronic and 
spintronic devices.

\section*{I-6  Doubling}

{\bf \huge Nielsen-Ninomiya's No-go theorem}

The authors are very proud  of, that we have shown a theorem 
saying:

{\bf
When one makes the mentioned  
``relativistic 
fermions of Weyl-type'' (=chirale fermion) on
a lattice (so e.g. in a crystal) then you  
always get equally many 
right-spinning and left-spinning Weyl-type particle(species). }

This theorem is a great challenge for those wanting 
to make a lattice model (with calculational purposes) 
for a theory with massless (or almost massless) quarks,
let alone the Standard Model.


{\bf By having 3 K +3 K$'$ Dirac-points of Compensating 
Handedness 
Our Doubling Theorem Realized in Graphene.}

\begin{center}
\includegraphics{perspectivegraphenebandstructure.pdf}

{\bf Our Doubling Theorem Realized in Graphene.}
\end{center}

\section*{I-7  ABJ Anomaly}
In the article

H. B. Nielsen and M. Ninomiya, ``Adler-Bell-Jackiw Anomaly 
And Weyl Fermions In Crystal,'' Phys. Lett. 130B, 389 (1983). 
doi:10.1016/0370-2693(83)91529-0

we have put forward how to understand intuitively the 
Adler-Bell-Jackiw anomaly and how it should be possible to 
see it in crystals. Indeed now it has -presumably- been found 
in $Na_3Sb$ in its three dimensional form; at least the 
characteristic property that this anomaly can lead to a 
negative magnetoresistance seems justified for this material as 
should be seen from the following figure:
\begin{center}
\includegraphics{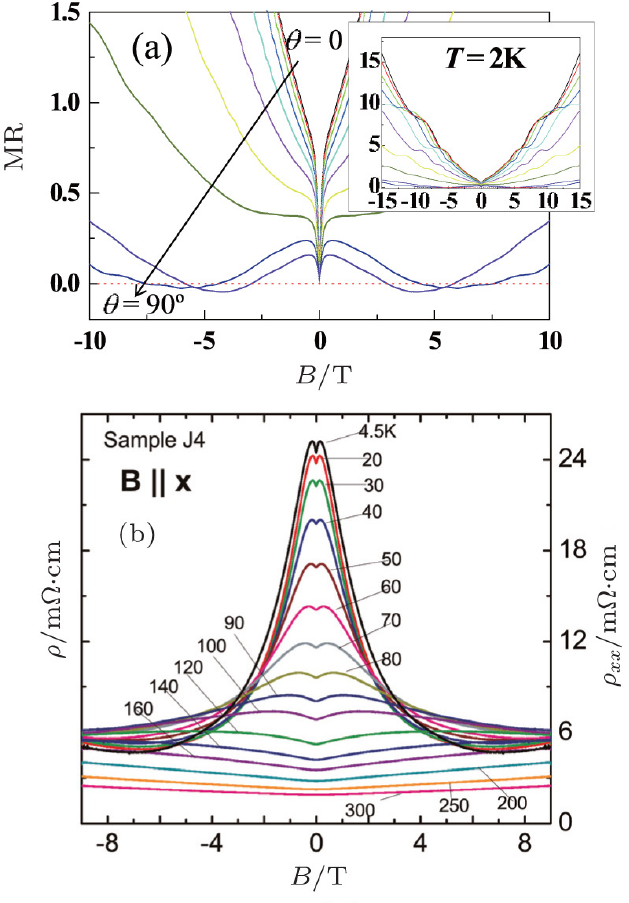}
\end{center}
It is clearly seen for the low temperatures that there is a 
dramatic peak in the resistance when the magnetic field is 
small, whereas the resistance becomes appreciably smaller 
when the magnetic field is switched on. The lower of the two 
figures shows the resistance in the direction of the magnetic 
field. It is indeed important that this increased 
conductivity goes in the direction of the magnetic field and 
thus there is a dependence of the magnetoresistance as a 
function also of the angle between the magnetic field and 
the direction of the electric field.

This subject will be explained in more detail in part II.

\section*{I-8  Further}

{\bf Further Developments of Our ``Random Dynamics''}
Further speculations, calculations, supporting the idea of 
getting the Standard Model out as a - say low energy limit -
of/from almost whatever the (most) ``fundamental'' physical 
laws (say complicated)  might be:

\begin {itemize}
\item A low energy boson system - with only momentum 
conservation ... like the general fermion system considered -
gives (in free approximation) free Maxwell equations.
\item Remarkably: All species of particles in the Standard 
Model {\bf except the Higgs boson} are eihter Yang-Mills 
particles or chiral fermions; so they would all be 
{\bf massless except for effects due to the Higgs field!}
This is just what one gets by asking for the low energy limit
in the general theory!  
\end{itemize}
\section*{I-9  Conclusion}

\begin{itemize}
\item 
Hope that the type of
{\color{red} relativistic 
chiral fermions},
one finds in high energy physics Standard Model
in fact  
{\color{red} comes by itself} -
and even points to the right dimensionality 
3 +1, 
which just is the right one-;
but there are  a couple of ``small'' problems
(different species of particles have in first go 
different ``maximal'' velocities)
\item Now adays the phenomenon is about being found in 
real materials, graphene etc. 
{\color{red}One can make relativity 
models chemically
}   
\end{itemize}

It should be especially stressed that {\bf the negative 
magneto-resistance due to the Adler Bell Jackiw anomaly has 
been seen in $Na_3Sb$.}

\newpage

\section*{II. What comes beyond 
Topological Insulator
?}
\vspace{-4mm}
--``Nielsen-Ninomiya Effect'' due to Adler-Bell Jackiw chiral Anomaly--
\section*{II-1 Introduction}
In part I we mainly argued about 
``Gapless Semiconductor'' 
``Topological Insulator'' and this subject 
has been very rapidly developing presently. 

 We now, in this part II, argue chiefly a 
new application of relativistic quantum 
field theory. Specifically, We investigate in 
condensed matter (in 
nano-scale$\cong 10^{-9}m$) how the 
Relativistic Quantum field theory Effect can 
appear and can be detected in material science.

Theoretically this effect was predicted 
already 35 years ago in 1983 by the present 
authors 
\begin{itemize}
\item
(H. B. N and M. N.) in a High Energy 
Theoretical Physics journal, Physics 
Letters B Vol. 130, issue 6 p 389 (1983), 
entitled ``The Adler-Bell-Jackiw anomaly and 
Weyl Fermions in a Crystal''. 
\item
Prior to the above paper one of the authors (M. N) 
was invited to give talks in the 
International Workshop on ``Lattice Field 
Theory'' in Saclay, Paris and Subsequently 
held XXI International Conference on High 
Energy Physics, Paris July 26-31, 1982 (so called ``Rochester Conference series''), where he 
talked about  Weyl fermions on lattices and the 
ABJ-anomaly.
\end{itemize}
In 
solid material there offen appears 
crystal lattice 
structure. Thus we are forced to use lattice 
field theory 
which has been well developped in high 
energy physics. In this formulation the 
crucial facts for us 
are the following:
 
Suppose At each lattice site we put one Weyl 
fermion e.g. $\Psi_L$ (Left-handed one).

Our Nielsen-Ninomiya Theorem states that 
there should appear 
equally many right handed and left handed 
Weyl fermions - looking in momentum space 
at different momentum values -. In the 
simplest construction resulting from just 
``naively'' replacing derivatives by 
differences on the lattice our theorem is 
implemented by there appearing 
$2^d$ species (d: space dimension). Therefore 
in 3 space dimensions it turs out that there 
should be 
8 species of Weyl (or chiral) 
fermions. Furthermore 
4 of them are 
left-handed $\Psi_L$ and rest 
4 species are 
right-handed $\Psi_R$ chiral fermions.

That is to say on the lattice there should 
be pairwise (left-handed and right-handed) 
chiral fermions. Therefore we are not able 
to construct chiral 
theory with for instance only one 
handed fermion on the 
lattice. Thus it leads to the very important 
consequence in high energy physics. In 
reality the Standard Model or, unified model 
of, weak and electromagnetic interactions 
called ``Glashow-Salam-Weinberg model'', 
or ``Standard Model'' of Weak and 
Electromagnetic Interaction cannot 
be constructed on the lattice! The reason is 
that in the Standard Model all the fermions 
are left-handed chiral fermions, while no right-handed fermion at all. The experimental 
results performed so far are all well in 
agreement with the standard model predictions.

If one takes serious the proposal of a new law 
of nature by one of us and various 
collaborators, ``Multiple Point Principle'',
one can even claim an indication for, that the 
Standard Model contrary to the expectation of 
many of our colleagues, should be valid up to 
an energy scale of the order of $10^{18} GeV$
(rather close to the Planck scale):

One of the authors (H. B. N.) 
made together with C. D. Froggatt  a 
theoretical calculation of $m_H$ with 
recourse to the just mentioned 
 ``multiple point principle 
(MPP)''. The value is in very good agreement 
with experimental value at LHC (Large Hadron 
Collider in CERN, Geneva) $m_H\sim$ 125GeV. 

See e.g. H. B. Nielsen and M. Ninomiya 
``Degenerate vacua from unification of 
second law of thermodynamics with other laws; 
The derivation of Multiple point principle'' 
Int. J. Mod. Phys. {\bf A23} (2008) 919 DOI: 10.1142/S02177510839682, in which an argument 
for among other things is given MPP from 
a model with the action taken to be complex 
rather than real as it is normal. 
\\

If the Standard Model shall as from this 
suggestion from Multiple Point Principle etc.
be valid only with tiny corrections if any 
almost up to the Planck scale, it would be 
even more mysterious that we could not put it 
on a lattice because of its chiral particles.
Really we could -it looks -hardly regularize 
it with any sensible cut off! Quite a mystery.
\cite{Rughmystery} 

2) ABJ anomaly on a lattice

Condensed matter researchers except for 
high energy physicists (including some 
nuclear theorists), may not have heard of the 
Adler-Bell-Jackiw or 
chiral anomaly.
Therefore we briefly explained ABJ anomaly in 
continuum space in Appendix A.

Here we turn to our nano-scale material case. 
In the material there is a lattice structure 
Fig. II-1
\begin{center}
\includegraphics[clip, width=10cm]{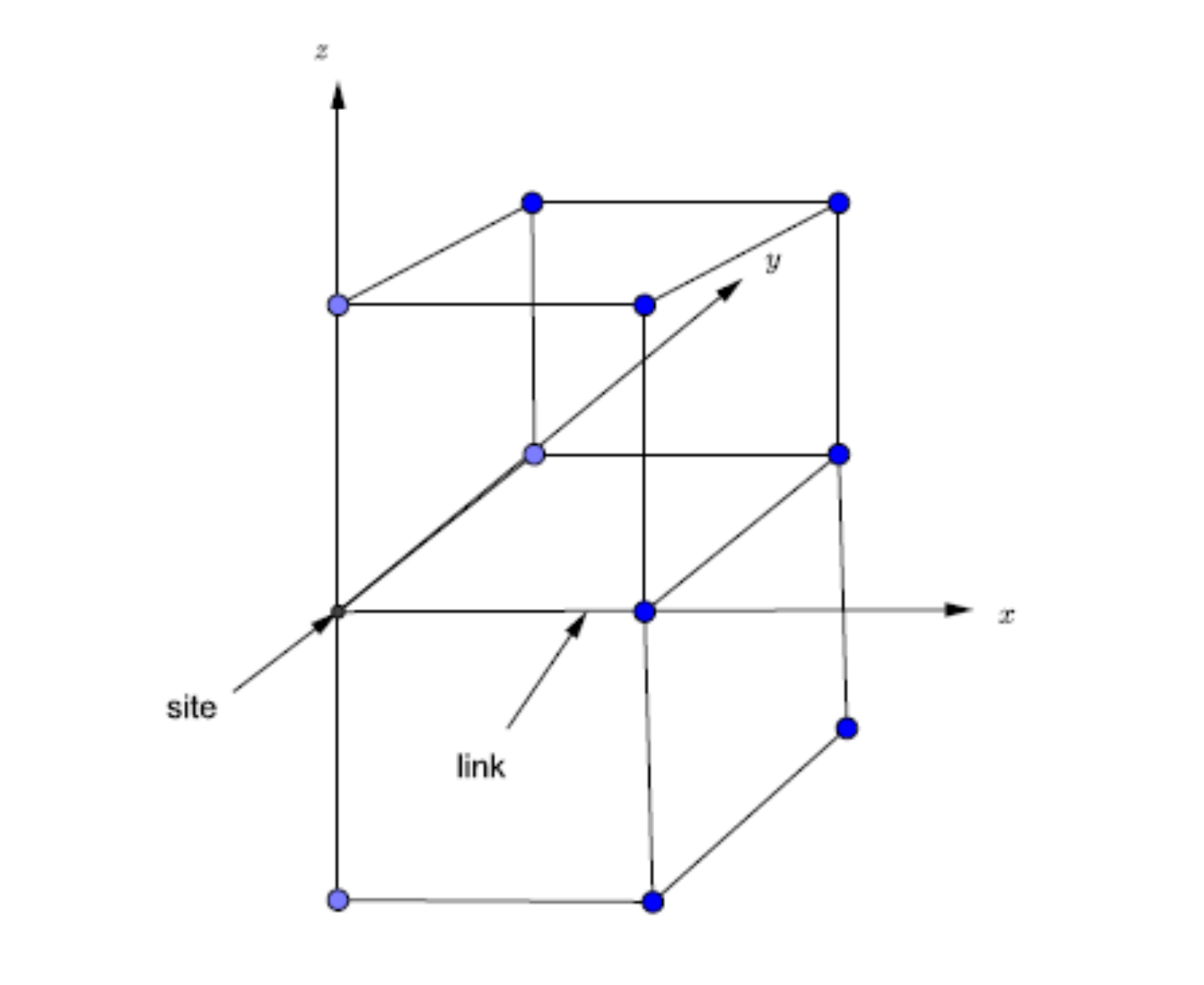}\\
{\bf Fig.}II-1
\end{center}

In this 3 dimensional lattice on each sites 
we put one Weyl or Chiral electron e.g. $e_L$ 
(Left handed electron),then  according to 
the Nielsen-Ninomiya Theorem, there should 
appear somehow so many of them, that there 
are equally many right haned and left handed 
ones. In fact we get in the simplest case
 4 $e_L$ as well as 4 $e_R$. 
\\

To understand band structure, we go to the 
momentum space. 

Note that due to the lattice translational 
invariance the momentum is conserved modulo 
multiple of the unit length of reciprocal 
lattice. 

The Brillouin zone in the momentum 
space is topologically equivalent to the 
hypertorus $S^1\times S^1\times S^1$. 

In such a topological structure of crystal 
lattice, the Adler-Bell-Jackiw anomaly 
explained 
for continuum spacetime in 
appendix B, is easily understood also , 
as was presented in PLB {\bf 130} n06, 
(1983) by the 
present authors.

\section*{II-2 $1+1$ dimensional 
example}
For simplicity, as an example the 1 space 
1 time 
dimensional case is considered.
Right chiral (Weyl) fermion obeys lattice 
Weyl eg. 

$i\frac{\partial}{\partial t}\Psi_R(na)=\frac{i}{2a}[\Psi_R((n+1)a)-\Psi_L((n-1)a)]$
where $n=0,\pm1,\pm2,\cdots$ denote sites and $a$ is a lattice space.
This can be easily solved and the dispersion relation is given by
$w=(\frac{1}{a})\sin pa$. Thus near $p=0$ there is a RH 
(RH = right handed) species 
with the dispersion law $w\approx p$ and 
further there is a 
 LH (LH =left handed) 
species near $\frac{\pi}{a}$ with the dispersion law 

$w\approx -(p -\frac{\pi}{a})$.

These situations are illustrated in the 
following Fig.

\begin{center}
\includegraphics[clip, width=15cm]{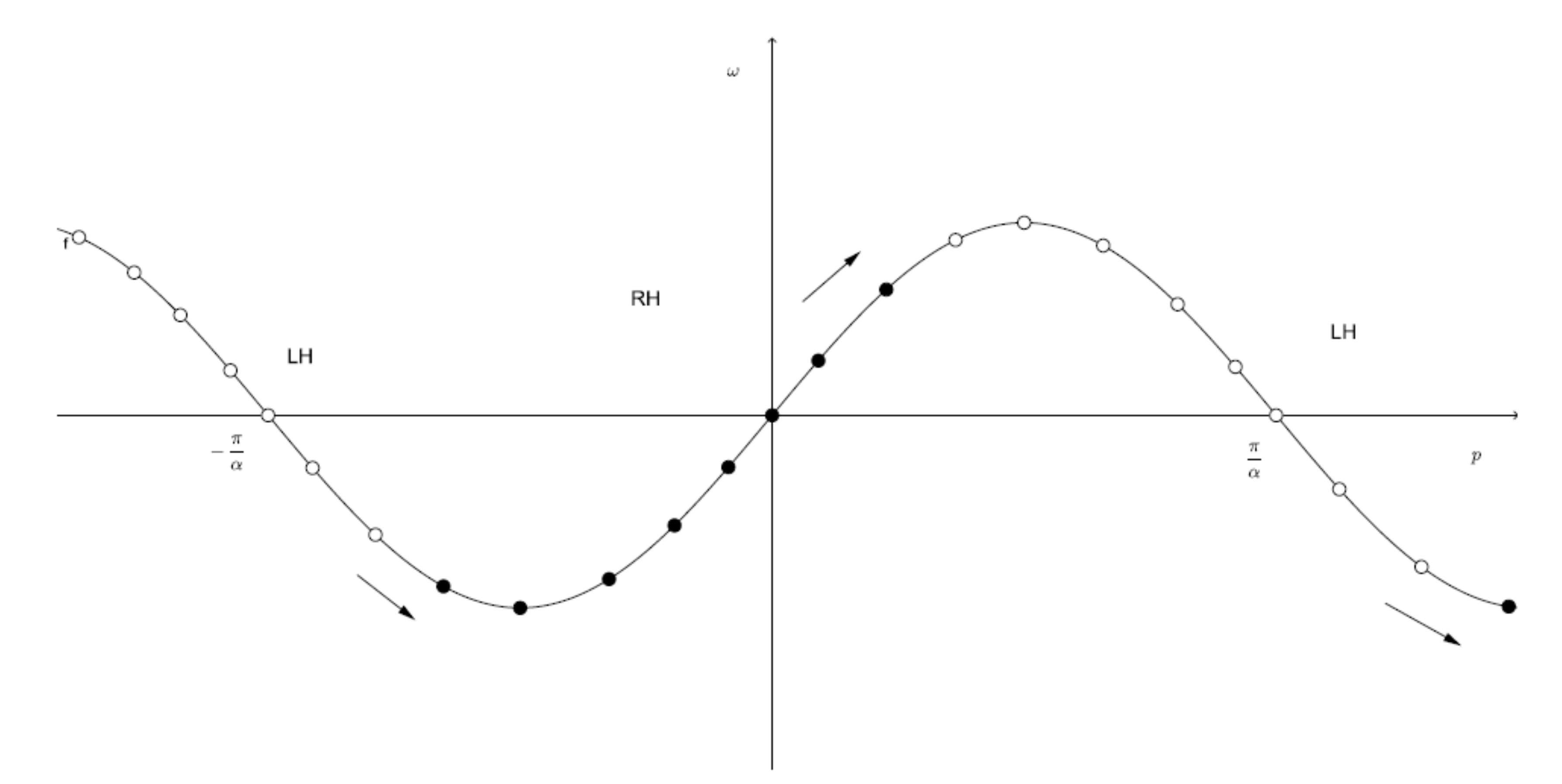}
\end{center}

Note that due to topology of momentum space, there is a 
periodicity modulo $2\pi$. (e.g. points $p=-\frac{\pi}{a}$ 
and $\frac{\pi}{a}$ are identified) 

\section*{II-3 $3+1$ dimensional case}
This $1+1$ dimension example clearly tells us, that in lattice 
theory there appear equal number of RH and LH chiral (or Weyl) 
fermion species (really in 1+1 dimension one should rather 
talk about right mover and left mover, because there is no 
genuine handedness in 1+1 dimensions) . It is not completely 
straightforward 
to generalize to $3+1$ dimensions, but with use of the 
appriate mathematics of homotopy (group) theory one 
make the analogous theorem in 3+1 or in even higher dimensions
to the theorem in 1+1 that in a period real function has 
pass zero in positive and in negative direction equally 
many times per period. 

\subsection*{II-3 (a) Weyl (or chiral) 
Fermion}
In generic chiral (Weyl) fermion theory which obeys

$i\dot{\Psi}(\overrightarrow{x})=
H\Psi(\overrightarrow{x})=w\Psi(\overrightarrow{x})$

We assume that the generic Hamiltonian satisfy the following four 
conditions:

(1)	Locality of interaction in the sense that 
$H(\overrightarrow{x}-\overrightarrow{y})\rightarrow 0$ 
as $|\overrightarrow{x}-\overrightarrow{y})|\rightarrow large$ 
fast enough that the Fourie transform of $H(\overrightarrow{x})$ 
has continuous first derivative.    
(2)	Translational invariance in the lattice
(3)	Hermiticiti of H (reality of S)
(4)	Furthermore an assumption is that the charge (=lepton number 
in our case) is bilinear in the fermion field. 

Under these 
conditions in the generic $H$ case we gave a rigorous proof in 
terms of the Homotopy theory in topology in 1981 (see, II-1).

\subsection*{II-3 (b) Adler-Bell-Jackiw 
anomaly on a lattice}
Let us go into the Adler-Bell-Jackiw (ABJ) anomaly on the 
lattice in the continuum spacetime. We reviewed this anomaly 
in continuum spacetime in Appendix B. 

Here we argue for the lattice version of the ABJ anomaly. Firstly we as 
as an example let us explain  the $1+1$ dimensional lattice Weyl 
(chiral) fermion. 
In the lattice RH chiral or Weyl electron system, we put on an 
external uniform electric field $E$  in $x$-direction denoted by 
$\dot{A}^1=E$ in temporal gauge $(A^0=0)$. Then the Weyl eq. 
reads

$i\frac{\partial}{\partial t}\Psi_R(x)=(-i\frac{\partial}{\partial x}-\dot{A}^1)\Psi_R(x)$.

The dispersion law is given by $\omega(p)=p$. 

In the classical eq. of the electron in the presence of the 
electric field is $\dot{p}=eE$ so that the RH electron in 
quantum theory is given by

$\dot{\omega}=\dot{p}=eE.$

Therefore the creation rate of the RH electrons per unit time 
and unit length is determined by a change of the Fermi surface 
that separates the filled and unifilled states as shown in 
Fig. II-2.

\includegraphics[clip, width=7cm]{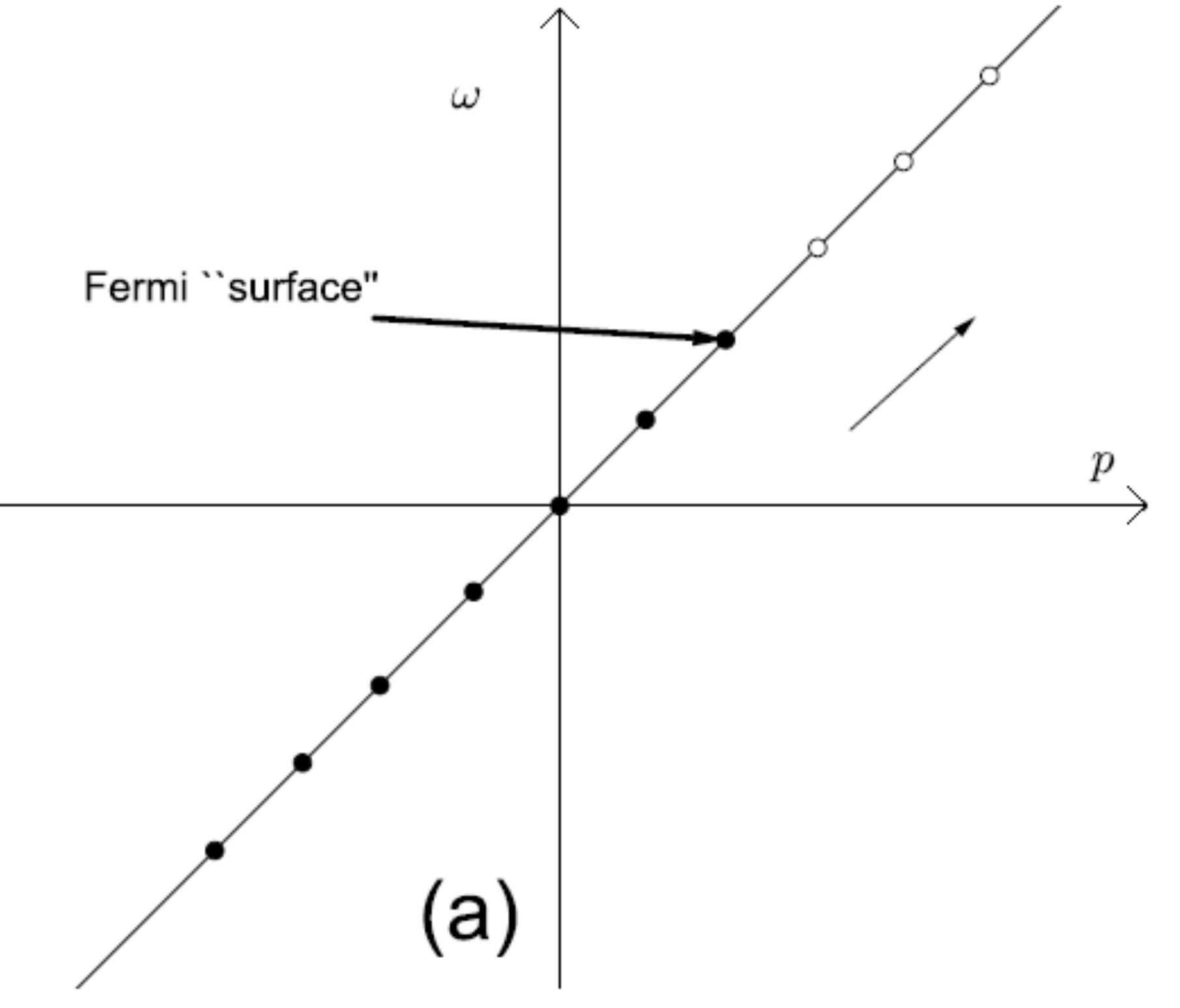}
\hspace{2cm}
\includegraphics[clip, width=7cm]{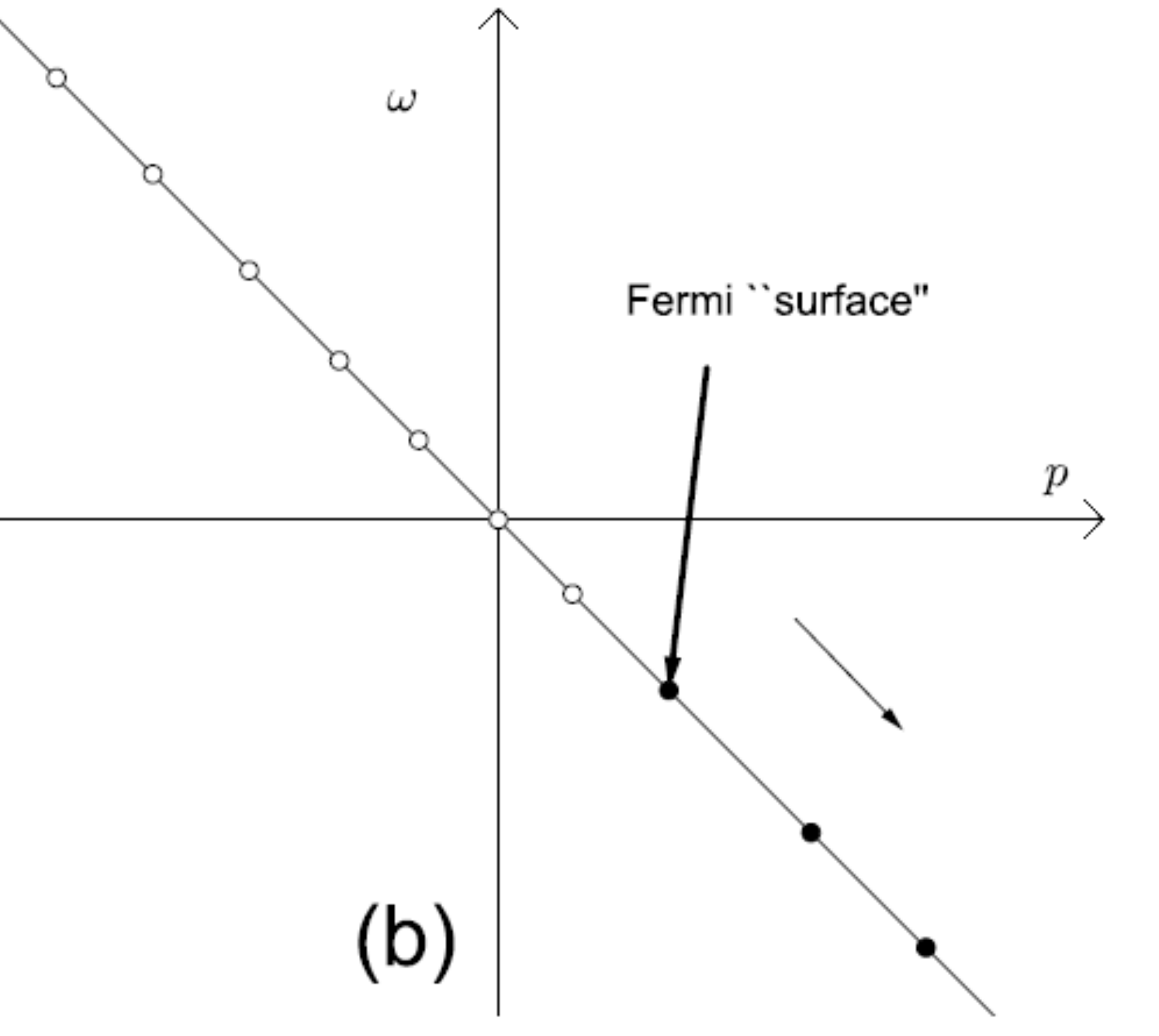}
\begin{center}
{\bf Fig.}II-2
\end{center}

We denotes the quantization length $L$, then the density of 
states per 
unit momentum is given by $\frac{L}{2\pi}$.
Therefore the rate of change of the RH electron number $N_R$ is 
given by $\dot{N_R}=
\frac{L}{2\pi}\cdot
\dot{\omega}_{fs}$

where $\dot{\omega}_{fs}$ denotes the 
rate of energy take up 
of the RH electron fermi surface per fermion, i.e. $eE$.

Therefore we obtain RH electron creation is given by 
$\dot{N}_R=\frac{e}{2\pi}E$ per unit length (namely for 
L=1).
This is the ABJ anomaly.

Thus the chiral charge $Q_R$ defined as the total 
number of RH particles (over the fermisea minus the 
number of holes) is not conserved: 
$\dot{Q}_R=\dot{{N}}_R=\frac{e}{2\pi}E$
In the same manner the annihilation rate of LH electrons 
with $\omega=-p$ is derived as $\dot{N}_L=-\frac{e}{2\pi}E$

This means that creation rate of the LH anti-electron is given 
as

$\dot{\bar{N}}_L=\frac{e}{2\pi}E$

By adding both, the anomaly of the Dirac electrons is

$\dot{N}_R+\dot{{N}}_L=\frac{e}{\pi}E$,
and thus

$\dot{Q}_5=\frac{e}{\pi}E$

To proceed to the $3+1$ dimension case, we should calculate 
the energy levels in the presence of an external uniform 
magnetic 
field, e.g. in the  
$z$-direction so that $A^2=Hx$, and $A^\mu=0$ otherwise. Thus we 
consider the equation for the two component RH electron 
field $\Psi_R$

$\left[i\frac{\partial}{\partial t}-(\overrightarrow{p}
-e\overrightarrow{A})\overrightarrow{\sigma}\right]\Psi_R(x)=0$

This eq. can be solved by introducing an auxiliary field 
$\Phi$ as

$\Psi_R=\left[i\frac{\partial}{\partial t}+(\overrightarrow{p}-e\overrightarrow{A})\overrightarrow{\sigma}\right]\Phi$.

Thus the eq. for $\Phi$ is given by

$\left[i\frac{\partial}{\partial}-(\overrightarrow{p}-e\overrightarrow{A})\overrightarrow{\sigma}\right]\cdot \left[i\frac{\partial}{\partial}+(\overrightarrow{p}-e\overrightarrow{A})\overrightarrow{\sigma}\right]\Phi =0$

This eq. reduces to the harmonic oscillation tpe eq,

$\left[-(\frac{\partial}{\partial x'})^2+(eH)^2(x'+\frac{p_2}{eH})+(p_3)^2+eH\sigma_3\right]\Phi=\omega^2\Phi$
with $\sigma_3=\pm1$

The energy eigenvalues $\omega$ are given by the Landau levels 
as follows
  
$\omega(n,\sigma_3,p_3)=\pm\left[2eH(n+\frac{1}{2})+(p_3)^2+(eH\sigma_3)\right]^{\frac{1}{2}}$  with $n=0,1,2,\cdots,$ except for the $n=0$ and $\sigma_3=-1$ mode. Here

$\omega(n=0, \sigma=-1,p_3)=\pm p_3$.

The eigenfunction is of the form

$\Phi_{n\sigma_3}(x)=N_{n\sigma_3}(x)\times{\rm exp}
(-ip_2x^2-ip_3x^3)\times{\rm exp}(-\frac{1}{2}eH
(x'+\frac{p_2}{eH})^2)\times H_m(x'+\frac{p_2}{eH})
\chi(\sigma_3)$

where $N_{n\sigma_3}$ is normalization constant and 
$\chi(\sigma_3)$ denotes the eigenfunctions of Pauli 
spin $\sigma_3: \chi(1)=\left(\begin{array}{c}1\\0\end{array}\right)$ and $\chi(-1)=\left(\begin{array}{c}0\\1\end{array}\right)$

Thus the solution of the eq. for Two-component RH 
electron $\Psi_R$ becomes the relations
$\Psi_R^{(n+1,\sigma_3=-1)}=\frac{N_{n+1},\sigma_3=-1}{N_{n,\sigma_3=1}}\Psi_R^{n,\sigma_3=1}$

for $n=0,1,2,\cdots$

The zero mode $n=0$ is

$\Psi_R^{(n=0,\sigma_3=-1)}=0$ with $\omega=-p_3$.

Therefore the ground state energy of $\Psi_R$ is given 
by $\omega(n=0,\sigma_3=-1,p_3)=p_3$ The 
energy 
eigenvalue for the other modes 
are

$\omega (n=0,\sigma_3,p_3)=
\pm\left[2eH(n+\frac{1}{2})+(p_3)^2+eH\sigma_3\right]^{\frac{1}{2}}$ 

These dispersion laws are depicted in the Fig.

\begin{center}
\includegraphics[clip, width=10cm]{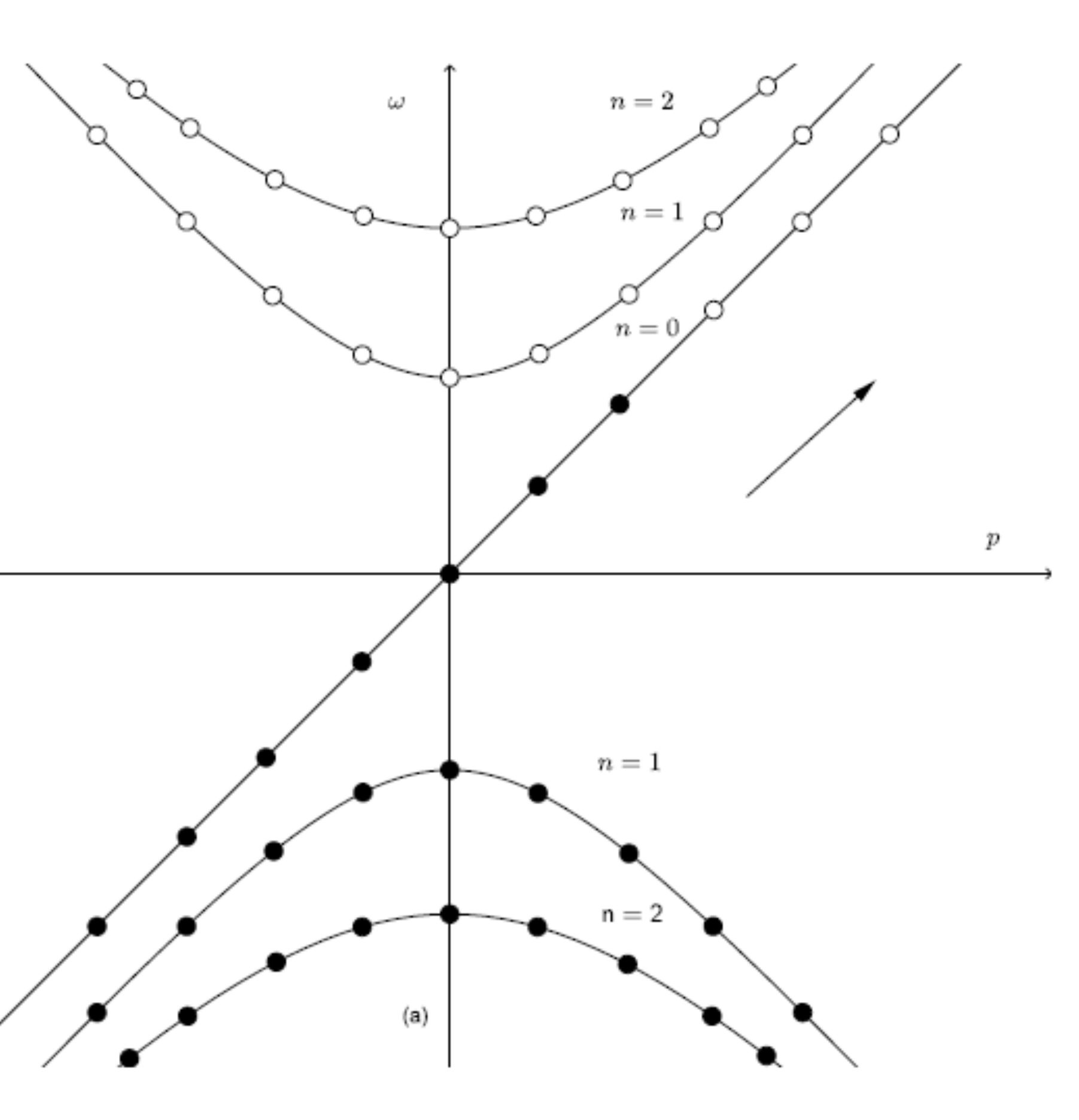}\\
{\bf Fig.}II-3
\end{center}

In the next step an external uniform electric field E is turned 
on along the same direction parallel to $H$. For the zero mode 
$(n=0,\sigma_3=-1)$ the dispersion law is the same as that 
for $1+1$ dimensions. Thus the creation rate of the particles 
is calculated in a similar manner. 

We should note that the electric field $E$ is switched on 
adiabatically,and  there is no particle creation in the 
$n\neq0$ modes. 
The density of the state in momentum space in the magnetic 
field direction is for quantization length $L$  
$L\frac{eH}{4\pi^2}$, 
and thus the creation rate (=the ABJ anomaly) is expressed as 
\begin{eqnarray*}
\dot{N}_R&=\frac{1}{L}\frac{LeH}{4\pi^2}\omega_{fs}\ \ (n=0, \sigma_3=-1, P_3)\\
&=\frac{e^2}{4\pi^2}EH\\
&=\dot{Q}_R
\end{eqnarray*}

For the LH electrons annihilation rate of LH anti electron is 
\begin{eqnarray*}
\dot{N}_L=-\frac{e^2}{4\pi}EH
\end{eqnarray*}
and the creation rate of the LH anti particle is given by
\begin{eqnarray*}
\dot{\bar{N}}_L&= \frac{e^2}{4\pi^2}EH\\
&=\dot{Q}_L
\end{eqnarray*}
In the case of the Dirac electron
\begin{eqnarray*}
\dot{N}_R+\dot{\bar{N}}_L&=\frac{e^2}{2\pi^2}EH\\
&=\dot{Q}_5
\end{eqnarray*}                                             
    
\subsection*{II-3 (c) Generic Case}

We again look at a generic case of which Hamiltonian is given 
by N$\times$N local Hermitian matrix. The N discrete energy 
eigenvalues are determined by the following eigenvalue eq. 
\begin{eqnarray*}
\sum^N_{l=1} H_{kl} (\overrightarrow{p}) \Psi_l^{(i)} (\overrightarrow{p})=\omega_i\Psi_k (\overrightarrow{p}) \ \ (i=1,\cdots,N)
\end{eqnarray*}
Here we assume that the $i$th level 
$\Psi_i(\overrightarrow{p})$ and $(i+1)$th level are degenerate. 
The eigenvalue $\omega_i(\overrightarrow{p})$ are assumed to be 
degenerate with the $(i+1)$ level at several different points
in momentum space,
 which are denoted as 
($\omega_d(\overrightarrow{p_d}), \overrightarrow{p}_d$) in the 
dispersion space ($\omega(\overrightarrow{p}),\overrightarrow{p}$). The $i$th and $(i+1)$th levels are described by d 
submatrix $H^{(2)}(\overrightarrow{p})$: it has the $i$th 
and $(i+1)$th entries of N$\times$N matrix H.

We then expand $H^{(2)}(\overrightarrow{p})$ in powers 
of $(\overrightarrow{p}-\overrightarrow{p}_d)$ around are of 
the degenerate point ($\omega_d(\overrightarrow{p}_d), p_d$). In 
the expansion of $H^{(2)}(\overrightarrow{p})$ is given
\begin{eqnarray*}
H^{(2)}(\overrightarrow{p})=H^{(2)}(\overrightarrow{p}_d)+(\overrightarrow{p}-\overrightarrow{p}_d)\frac{\partial H^{(2)}(\overrightarrow{p})}{\partial\overrightarrow{p}}|_{\overrightarrow{p}=\overrightarrow{p}_d}+O((\overrightarrow{p}-\overrightarrow{p}_d)^2).
\end{eqnarray*}

The derivative term is expressed by the Pauli matrices $(\_\hspace{-2.5mm}1+\sigma_\alpha)$, $(\alpha=1, 2, 3)$ and $\_\hspace{-2.5mm}1=2\times2$
unit matrix, as 
\begin{eqnarray*}
\frac{\partial H^{(2)}}{\partial\overrightarrow{p}_k}|_{\overrightarrow{p}=\overrightarrow{p}_d}=a_k(\overrightarrow{p}_d)\_\hspace{-2.5mm}1+V^\alpha_k(\overrightarrow{p}_d)\sigma_\alpha
\end{eqnarray*}
Here V are the constants depending on $\overrightarrow{p}_d$. 
Thus near $\overrightarrow{p}=\overrightarrow{p}_d$, 
$H^{(2)}(\overrightarrow{p})$ takes the form 
\begin{eqnarray*}
H^{(2)}(\overrightarrow{p})=\omega_d\_\hspace{-2.5mm}1+(\overrightarrow{p}-\overrightarrow{p}_d)\overrightarrow{a}\_\hspace{-2.5mm}1+(\overrightarrow{p}-\overrightarrow{p}_d)_k V^k_\alpha\sigma^\alpha
 \end{eqnarray*}

The eigenvalue eq. of the $i$th and $(1+i)$th energy 
eigenvalues near $\overrightarrow{p}=\overrightarrow{p}_d$ 
$H^{(2)}(\overrightarrow{p})u=\omega u$.

This is rewritten by using a new set of variables 
\begin{eqnarray*}
\hat{p}=\overrightarrow{p}-\overrightarrow{p}_d, \ \ p^0=\omega-\omega_d-\hat{p}\overrightarrow{a}
\end{eqnarray*}
as
\begin{eqnarray*}
\hat{p}V\overrightarrow{\sigma} \ \ u=p^0u
\end{eqnarray*}
If we introduce
\begin{eqnarray*}
K^0=p^0\ \ {\rm and}\ \ k=\pm \hat{p}\overrightarrow{V}
\end{eqnarray*}
Were $\pm$ correspond to the sign of det $V$.
For simplicity we may take as an example $V_{k\alpha}=v\delta_{k\alpha}$ $(k,\alpha=1,2,3)$.

The above eigenvalue eq. becomes 
\begin{eqnarray*}
\overrightarrow{k}\overrightarrow{\sigma}u=\pm k^0u
\end{eqnarray*}
Where the dispersion law $(k^0)^2=v^2k^2$. Thus, it 
is $\omega^2=v^2p^2$ 

In this way RH and LH Weyl eq. describes the 2 energy levels 
near degeneracy point in 
$(\omega(\overrightarrow{p}),\overrightarrow{p})$ space 
correspond to a species of Weyl fermions contained in the 
theory. Our theorem tells that RH and LH degeneracy points 
appear necessarily as a pair because of the Brillouin zero 
structure (topology). The theorem was proved by only 
topological arguments together with locality, as was shown 
our papers in 1981. The doubling of the Weyl fermions are 
illustrated in Fig. II-4 (page 18).


\section*{II-4 Parity non-invariant zero-gap material}\label{Paritynon}
We assume that we have found a parity non 
invariant 
material (i.e. a crystal should be of 
non-centrosymmetric symmetry; e.g. BiTeI form 
a non-centrosymmetric crystal.
Best might be a triclinic pedial class with 
no point symmetry at all.) with zero-gap,
 which can be simulated by a Weyl, 
fermion theory with 
a dispersion law $\omega^2=v^2p^2$. The 
effect analogous to 
the ABJ anomaly gives rise to a peculiar 
behavior of the 
conductivity of the electric current in the 
presence of the 
magnetic field. It is enough to consider one 
conduction 
band $\omega_i$. 

 The valence band $\omega_{i+1}$ (negative 
energy state) is 
assumed to be completely filled. In the 
absence of external 
field, the single electron distribution 
function in the 
thermodynamical equilibrium is of the form
$f_0(\overrightarrow{p})=[1+{\rm exp}[(\omega(p)-u)/kT]]^{-1}$

In the presence of $E$ and $H=0$ there occurs 
a small deviation from 
thermodynamical equilibrium
so that $f=f_0+\delta f$, and  the $E$ field 
accelerates the electrons in the same 
direction and then 
\begin{eqnarray*}
\left(\frac{\partial f}{\partial t}\right)_{\rm drift}=eE\frac{\partial f}{\partial p_z}.
\end{eqnarray*}
At the same time the accelerated electrons 
get scattered back into some states in the 
same cone. We assume that f fills back into 
$f_0$ exponentially with a relaxation time 
$\tau_0$ so that 
$\delta f \propto e^{-\frac{\tau}{\tau_0}}$

Then 
\begin{eqnarray*}
\left(\frac{\partial f}{\partial t}\right)_{\rm coll}=-\frac{1}{\tau_0}(f-f_0)
\end{eqnarray*}
Therefore the steady state condition is $\left(\frac{\partial f}{\partial t}\right)_{\rm drift}=-\left(\frac{\partial f}{\partial t}_{\rm coll}\right)$
(Boltzmann eq.).

The sol. of this is in the lowest order in $E$ 
\begin{eqnarray*}
f(\overrightarrow{p})=f_0(\omega)+eE\tau_0\frac{\partial f(\omega)}{\partial p_z}
\end{eqnarray*}
Then the longitudinal current density is given by
\begin{eqnarray*}
J_0=\frac{1}{L^3}\sum_{\overrightarrow{p}}(-e)v_z f(\overrightarrow{p})({\rm \#deg.\ pts})
\end{eqnarray*}
Where $v_z=\frac{\partial \omega}{\partial p_z}$
and $({\rm \#deg.\ pt})$ denotes the number 
of deg. pts (= degeneracy points).

In the low temperature approximation 
$f_0(\omega)=\theta(\mu-\omega)$ so that
\begin{eqnarray*}
J_0=\frac{1}{6\pi^2}e^2E(\frac{\mu^2}{v})\tau_0({\rm \#deg.\ pt})
\end{eqnarray*}
the relaxation time is given in terms of 
transition probability of electron from the 
state with 
$\_\hspace{-2.5mm}\overrightarrow{p}$   
  into one with 
$\_\hspace{-2.5mm}\overrightarrow{p}'$,
 $W(\overrightarrow{p}\rightarrow\overrightarrow{p}')$
by
\begin{eqnarray*}
\frac{1}{\tau_0}=\frac{1}{L^3}\sum_{\overrightarrow{p}'}\frac{p_z-p_Z'}{p_z} W(\overrightarrow{p}\rightarrow\overrightarrow{p}')
\end{eqnarray*}
We assume that the interaction between the 
electron and the ionized impurities is given 
by the screened Coulomb potential (pot.) of 
the from 
\begin{eqnarray*}
V(\overrightarrow{x})=\left(\frac{4\pi e^2}{k}\right)\frac{e^{-\frac{|\overrightarrow{x}|}{\gamma_0}}}{|\overrightarrow{x}|}
\end{eqnarray*}
With the screening length $\gamma_0$ and $k$ 
the dielectric constant. Computing  $\tau_0$ 
in the first order perturbation we obtain the 
current as
\begin{eqnarray*}
J_0=\frac{4e^2E}{3\pi\eta_I}\left(\frac{k}{4\pi e^2}\right)^2\left(\frac{\mu^4}{v^2}\right)\left[{\rm ln}(1+\beta)-\frac{\beta}{1+\beta}\right]^{-1}({\rm \# deg.\ pt})
\end{eqnarray*}
With $\beta=\frac{2\pi k v}{e^2}({\rm \# deg.\ pt})$
and $\eta_I$ the density of impurity. 

Next compute the magneto-conductivity 
when H parallel to E is so strong that only 
the lowest states $n=0, \sigma_3=-1$
with dispersion law $\omega=vp_z$ or 
$\omega=-vp_z$
near the RH and LH degeneracy point are 
filled the ABJ anomaly effect will cause 
the movement 
in the momentum space of electrons from the 
lowest Landau level $(n=0, \sigma_3=-1)$
at the one deg. pt. (=degeneracy point) in 
the LH cone to the 
corresponding one $(n=0, \sigma_3=-1)$ 
in the RH cone (at the RH deg.pt.). Thus 
these moved electrons  will 
give raise to a deviation from the 
thermodynamical equilibrium, that can be 
expressed by the different chemical 
potentials for the electrons at the RH 
degeneracy pt., $\mu_R$ and at the LH 
one $\mu_L$.
If one had calculated the relaxation time in 
the 
approximation where only one degeneracy point 
at a time was relevant -such as we did above  
in the $H=0$ case--we would have found $\frac{1}{\tau}=0$. This comes out of such a 
calculation due to the energy conservation 
factor $\delta(\omega-\omega')=\frac{1}{v}\delta(p_z-p_z')$
contained in $W(P_Z-P_Z')$ which makes (23) 
give $\frac{1}{\tau}=0$. However we cannot 
neglect scattering processes involving two 
degeneracy point. 

\section*{II-5 Transfer from LH to RH cones by Adler-Bell-Jackiw Anomaly}
The mechanism for the electric current with 
both $E$ can $H$ 
switched on 
peculiarly different from the one with a 
negligibly weak $H$. In the presence of 
strong $H$ the lattice anomaly of the ABJ 
anomaly takes place: transfer of the 
particles from the LH degeneracy pt. to the 
RH one acts as a drift term, i.e. 
$\dot{N}|_{\rm drift}$ in the Boltzmann 
equation. On the other hand for 
negligible $H$ each degeneracy points act 
independently. By the ABJ anomaly the Fermi 
energy level $\mu_R$ in the RH cone goes up 
compared to that of the $H=0$ case $\mu$ 
and $\mu_L$ in the LH cone is lowered. See 
Fig.  II-3 (a) ($1+1$ dim. case) and Fig. II-4 ($3+1$ dim. case)
\begin{center}
\includegraphics[clip, width=10cm]{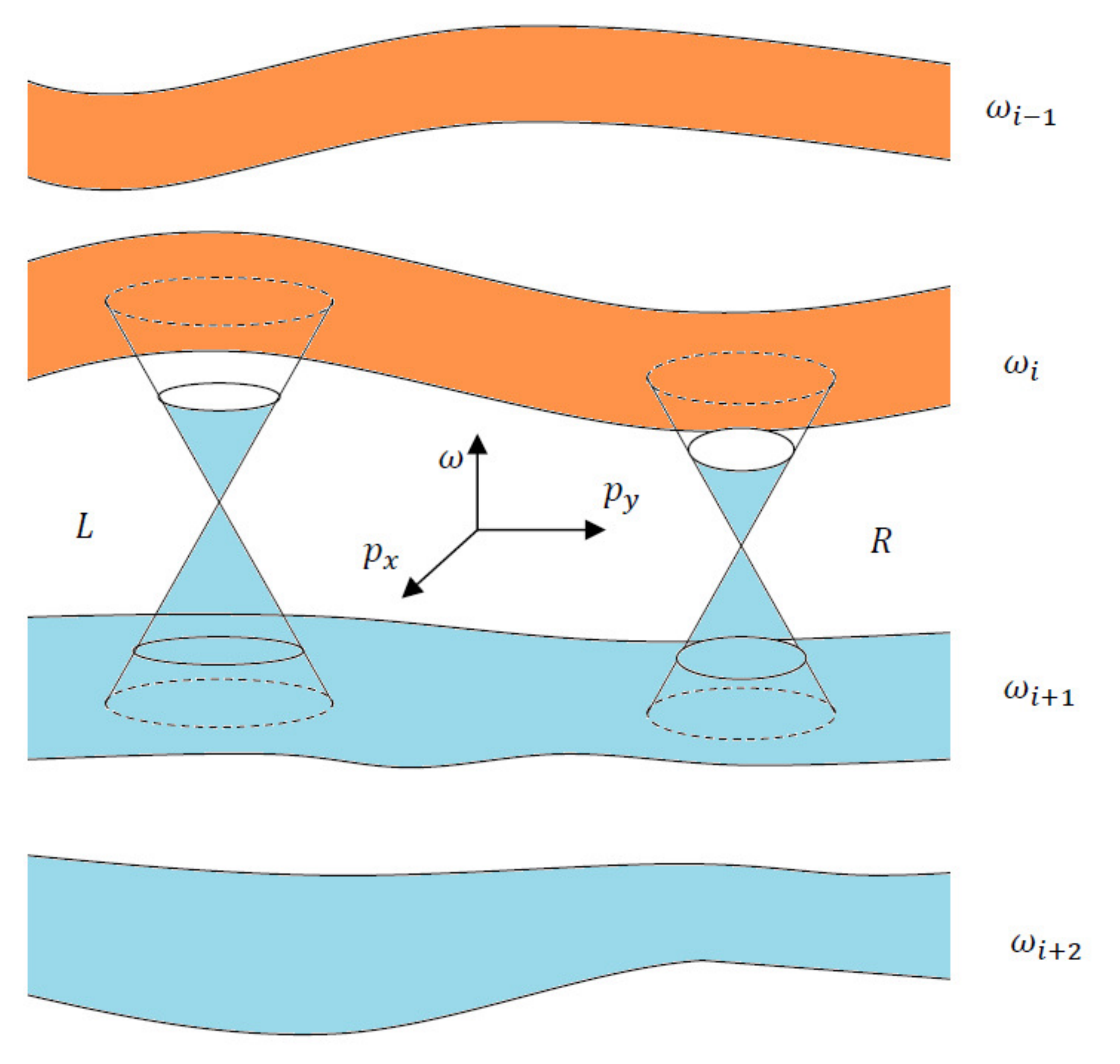}\\
{\bf Fig. }II-4
\end{center}

In order that the system is in the steady 
state

{\bf the excess electrons by the ABJ anomaly 
in the RH cone must be scattered back to the 
another state}.

But they can not be scattered back into the 
state in the same come! because, as was 
explained above $\tau=\infty$.

Therefore they must transfer into the 
states in another cone; that is from the RH 
cone into the LH cone. 

We may call this 
the intercone scattering and we denote the 
corresponding relaxation time by $\tau_I$. If 
the intercone transition probability $W(p_z\rightarrow p_z')$ from RH cone into the LH cone 
is calculated, then the collision term is 
given by 

\begin{eqnarray*}\dot{N}_R|_{\rm coll}&=\frac{2}{L}\sum_{p_Z}[f(p_Z)-f_0(p_z)]\frac{1}{L}\sum_{p_z'}W(p_z\rightarrow p_z')\\
&\equiv -\frac{p_z'}{\tau_I}(N_R-N_R^0)
\end{eqnarray*}

Here $N_R$ and $N_R^0$ denote the total 
electron numbers in the RH cone above the 
degeneracy energy in the $H\neq0$ and $H=0$ 
cases respectively.
Thus 
$\frac{1}{\tau_I}=\frac{2eH}{(2\pi)^2}\frac{1}{L}\sum_{p_3'}W(p_z-p_z')$

The generation of a current associated with 
the ABJ anomaly can be shown by the 
following energy conservation argument. ABJ 
anomaly indicates that electrons are 
transferred from the LH cone into the RH cone 
by the rate of $\frac{e^2EH}{(2\pi)^2}$ per 
unit Time,per unit volume.: Notice that the 
dispersion law is continuous and the RH and LH cones are connected smoothly as shown Fig. 
II-4

Since the Fermi level energies are 
$\mu_R>\mu_L$ the transfer costs the 
energy $\frac{e^2}{(2\pi)^2}EH(\mu_R-\mu_L)$. 
This energy must be taken from the E field 
by the presence of a current $J_A$ determined 
by the energy balance as 

$EJ_A=\frac{e^2}{(2\pi)^2}eH(\mu_R-\mu_L)$

At the zero temperature, in the RH cone

$f_0(\omega)=\theta(\mu_R-\omega)$ and
thus
\begin{eqnarray*}
N_R&=\frac{1}{L^3}\sum_{p_yp_z}f_0(\omega)\frac{eH}{(2\pi)^2}\frac{\mu_R}{v}\\
&\cong N_R^0+(\mu_R-\mu)\frac{\partial N_R}{\partial \mu}
\end{eqnarray*}

Inserting this into Boltzmann eq. 

$\dot{N}_R|_{\rm drift}=-\dot{N}_R|_{\rm col}$

We obtain 
$\mu_R-\mu_L=evE\tau_I$

Therefore 
$J_A=ev\frac{e^2}{(2\pi)^2}EH\tau_I({\rm \# deg. pt.})$
Here the subscript A stands for the 
anomalous current the one associated with 
the analogue of the ABJ anomaly. In the 
definition of $\tau_I$ we may approximate 
$W(\overrightarrow{p_z}\rightarrow p_z')\cong W(\overrightarrow{p}-\overrightarrow{p'})$

So that 
$W(p_Z-p_Z')\cong(\frac{4\pi^2}{k})^2\eta_I
\left[(\overrightarrow{p}-
\overrightarrow{p}')^2+\frac{1}{\gamma_H^2}
\right]^{-2}2\pi\delta(\omega-\omega ')$
with $\frac{1}{\gamma_H^2}=\frac{EH}{kv}
({\rm \# deg. pt.})$.
According to 
$\hat{\overrightarrow{p}}\equiv\overrightarrow{p}-\overrightarrow{p}_d$, $p^0=\omega-\omega_d-\hat{p}\overrightarrow{a}$,
we have
$\overrightarrow{p}-\overrightarrow{p}'=\overrightarrow{p}_d-\overrightarrow{p}_d'+\hat{\overrightarrow{p}}-\hat{p}'$

where $\hat{\overrightarrow{p}}$ and $\hat{\overrightarrow{p}}'$ are oscillating around $\overrightarrow{p}_d$ and $\overrightarrow{p}_d'$: since they are order of $(eH)^{\frac{1}{2}}$. We may ignore the oscillatory part $(\hat{\overrightarrow{p}}-\hat{\overrightarrow{p}}')$ and $\frac{1}{\gamma_H^2}$ term in the denominator of $W(p_Z-p_Z')$
when compared to the distance of the RH and LH deg. pts $\overrightarrow{p}_d-\overrightarrow{p}_d'$.
In this approximation we obtain 

$J_A=\frac{e^2v^2E}{2\pi\eta_I}\left(\frac{k}{4\pi^2}\right)^2(\overrightarrow{p}_d-\overrightarrow{p}'_d)^4({\rm \# deg. pt.})$

We then obtain the ratio of the conductivity that is defined by $f=\sigma E$ as $\frac{\sigma_A}{\sigma_0}=\frac{3}{16}\left(\frac{v}{\mu}\right)^4\left[{\rm ln}(1+\beta)-\frac{\beta}{1+\beta}\right](\overrightarrow{p}_d-\overrightarrow{p}_d')^4$

By these results, for the intercom relation 
time $\tau_I$ the electrons must travel 
a ``long distance'' in momentum space. Thus 
$\tau_I$ is expected to be a large value 
compared to $\tau_0$ for $H=0$. Therefore 
$\frac{\sigma_A}{\sigma_0}$ given above is 
large.  

\section*{II-6 Further arguments}
So far we have presented our own theoretical 
predictions in 1983 although we believed 
sooner or later our predicted 
``Nielsen-Ninomiya'' mechanism (or effect) 
will be proved by experiment. Indeed after 
almost 35 year later Princeton University 
group led by Prof. N. Phuan Ong and R. Cava, 
found chiral anomaly in crystalline material. 
This surprising news in science community 
appeared in an article by Catherine Zandonella,
\begin{itemize}
\item office of the Dean of Research, in 
Science, September 3, 2015 entitled            
Research at Princeton: 
Long-sought chiral anomaly detected in 
crystalline material (science). \\

At the almost same time, scientist's article 
entitled.
\item ``Evidence for the chiral anomaly in 
the Dirac semimetal Na3Bi'' 
By J. Xiong Satya K. Kushwaha, Tian Liang, J. W. Kritzan, M. Hirsehberger, Wulin Wang, R. J. Cava, X. P. Oug, Science Express, 03 , September 2015.
and
 \item ``Signature of the chiral anomaly in 
a Dirac semimetal -- a current plume steered 
by J. Xiong, S. K. Kushwaraha, T. Liang, J. W. Krizan, Wudi Wang, R. J. Cava and N. P. Ong
\end{itemize}

Since then the works on this subject is 
really under rapidly developing mainly in 
Experiments, also theories: e.
g. 
 Dirac cones,and Weyl semimetals. 
We believe in the rather near future we 
shall see some 
machines using ``Nielsen-Ninomiya Mechanism (or Effect).
See e.g also \cite{ZnTe4}.
\section*{II-7 Conclusions}
In the present article we present the 
viewpoint at two exceptional high energy 
theoretical physicists new eras of condensed 
matter. 

In the first part I we mainly considered 
 ``Topological 
Insulator'' from random dynamics point of 
view. 
The essential point is that in generic 
Fermion dispersion relations i.e. in (almost) 
all 
solids or fluids at low temperature  we can 
derive the recently found properties of 
Topological insulators such as graphene etc. 

In the 2nd point II, we present what comes 
beyond topological insulator.

We believe that the Adler-Bell-Jackiw anomaly 
effect in the chiral non invariant gapless 
material, causes that
\begin{itemize}
\item magnetic conductance is enhanced very 
much (ideally permanent current)
\item Chiral electron (chiral fermion in 
general) in lattice of the gapless material 
runs with a fixed speed. (This fixed speed 
is what in the relativity theory analogue 
is the speed of light.)
This is so, because we by analogy can apply 
the 
relativistic quantum 
field theory.
\end{itemize}
To make any apparatus using the above theory 
will be widely opened to not only condensed 
matter, but chemistry, beyond artificial 
division such as, physics chemistry 
engineering etc.        

\newpage
\section*{Acknowledgements}
One of the authors (H. B. N.) acknowledges 
the Niels Bohr Institute, Copenhagen 
University for allowance to work as emeritus
and for economical support to visit the 
conference in New York for which this work 
is the 
proceeding. 
M. N. acknowledges Advanced Mathematical 
Institute Osaka City University, and Yukawa 
Institute for Theoretical Physics, Kyoto 
University as emeritus. 

M. N. also acknowledges the present research 
is supported by the JSPS Grant in Aid for 
Scientific Research No. ISKO 5063.

\newpage

\section*{Appendix A}
We consider electron in quantum field 
theory (Relativistic quantum mechanics.) We present only necessary properties in Appendix A

The electron in the relativistic quantum 
field theory it is usually described as 
Dirac field 
$\Psi=\left(\begin{array}{c}\Psi_L\\\Psi_R\end{array}\right)$    
where $\Psi_L$ and $\Psi_R$ are 2 component 
fields. Now the electron has intrinsic spin
$\overrightarrow{S}$. 
Thus electron has the angular momentum then $\overrightarrow{J}$, 
whose value are half integers, and 
the spin components 
is $\overrightarrow{S}$ take values $\pm \frac{1}{2}$.
 

For massless fermions the right $\Psi_R$ and 
the left $\Psi_L$ componets in the (free) Dirac equation gets 
seperated, and we actually even find that the  
spin direction is the same as that of electron movement for the right components $\Psi_R$ and the opposite for the left components 
$\Psi_L$. Let us start with the Dirac field 
such as an electron in the quantum field 
theory. 
The electron has intrinsic spin $\frac{1}{2}$ 
of fermion obeying the free Dirac eq. 

$(i\gamma^\mu\partial_\mu-m_e)\Psi_D=0\ \ \ \ ({\rm II}-1)$ 
\\
thereafter we ignore electron mass unless 
described. Our notation is that of the 
textbook of Bjorken-Drell `` Relativistic 
Quantum Fields''. For our purpose we list up 
relevant notations below 
\begin{itemize}
\item The $3+1$ dimensional flat space 
metric (tensor):

$g_{\mu\nu}=\left(\begin{array}{cccc}
1&0&0&0\\
0&-1&0&0\\
0&0&-1&0\\
0&0&0&-1\end{array}\right)$

\item The $\gamma$ matrices are

$\gamma^0=\left(\begin{array}{cc}\_\hspace{-2.5mm}{1}&0\\0&-\_\hspace{-2.5mm}{1}\end{array}\right)$

were \ $\_\hspace{-2.5mm}{1}=\left(\begin{array}{cc}1&0\\0&1\end{array}\right)$

and \ $\gamma^i=\left(\begin{array}{cc}0&\sigma^i\\-\sigma^i&0\end{array}\right)$ $i=1,2,3$

Here $\sigma^i$ denotes $2\times 2$ Pauli matrices and  $\_\hspace{-2.5mm}{1}=\left(\begin{array}{cc}1&0\\0&1\end{array}\right)$.

Furthermore

$\gamma^5=\gamma_5=\left(\begin{array}{cc}0&\_\hspace{-2.5mm}{1}\\\_\hspace{-2.5mm}{1}&0\end{array}\right)$  (note $ \left(\gamma^5\right)^2=1$)
\item The 4 component Dirac field is denoted as

$\Psi_D(p,s)$

and when there is no interactions obeys the free Dirac eq. as 

$(i\gamma^\mu\partial_\mu-m)\Psi=0$

where $\partial_\mu=\frac{\partial}{\partial x^\mu}$, in momentum representation

$p^\mu=i\frac{\partial}{\partial x_\mu}$ 

$\Psi_D(\overrightarrow{p},s)$.

Here s denote intrinsic spin $|s|=\frac{1}{2}$

$\Psi_D(\overrightarrow{p},\overrightarrow{s})$ obeys
 
$/\hspace{-2.2mm}p\Psi_D(p,s)=0$

with $4$ component Dirac field we may describe 

$\Psi_D(p,s)=\left(\begin{array}{c}\Psi_L\\\Psi_R\end{array}\right)(p,s)$

Where $\Psi_L$ and $\Psi_R$ are $2$ component 
spinor respectively the eigenvalue solution of free Dirac eq. $({\rm II}-1)$
is the form of

$\Psi(p,s)=\pm \sqrt{\overrightarrow{p}^2+m^2}$

\end{itemize}
\begin{center}
\includegraphics[clip, width=10cm]{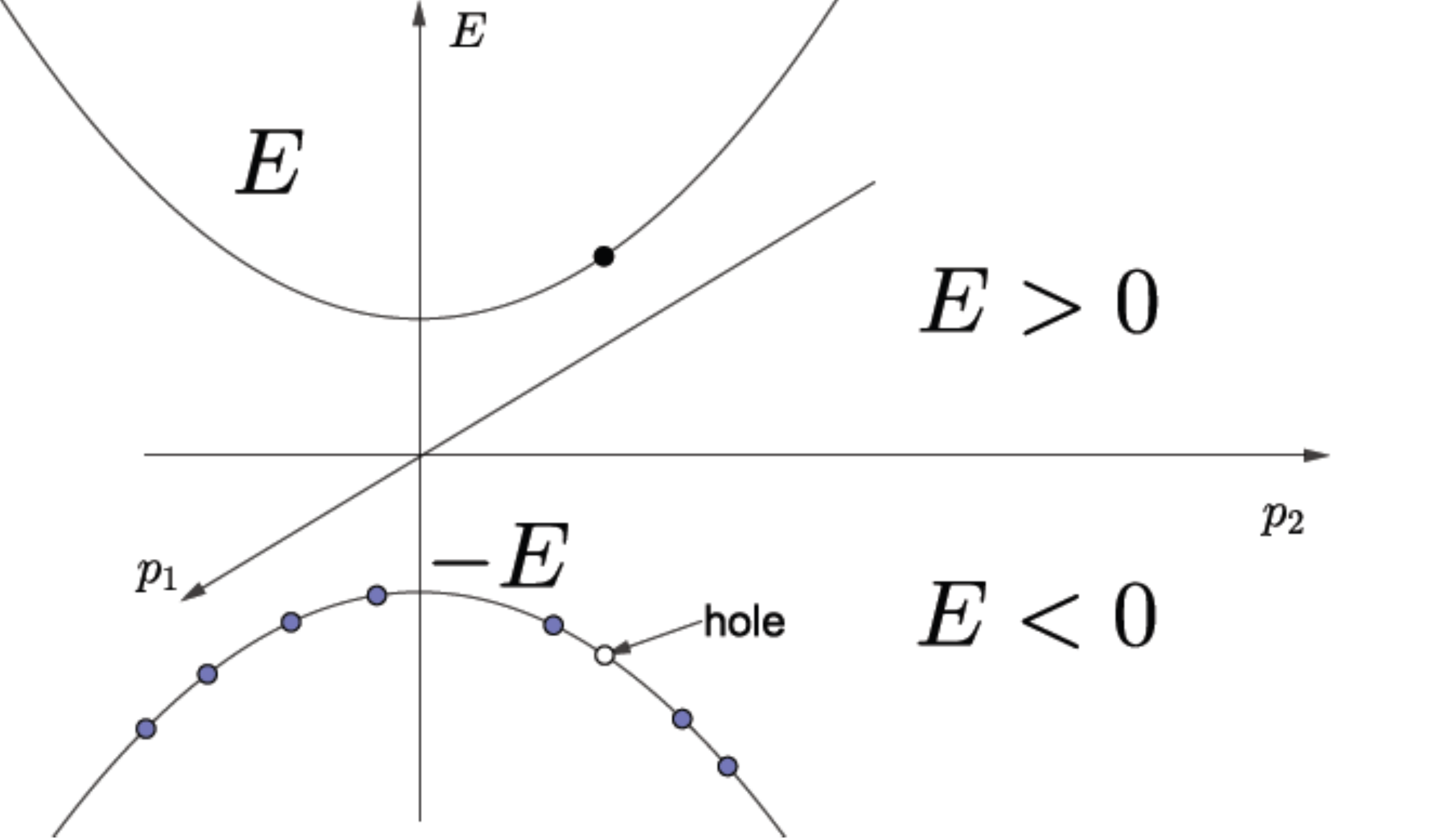}
\end{center}

We adopt the Dirac's ``hole theory''. In this 
theory often used in condensed matter as 
dispersion relation, the negative states are 
all filled, while the hole in the Dirac sea 
is antiparticle, i.e. positron $e^+$.

In solid state physics where one has say a 
crystal lattice, which from the quantum field
theory is discretized, so therefore we are 
interested in discretizing the quantum field 
theory here.
The  Dirac fermion wave function $\Psi_D(\overrightarrow{p},s)$ has $4$-components:
2 degree of freedom as that energy can have plus or minus. Furthermore the electron has 
intrinsic spin of which value is $|s|=\frac{1}{2}$. In the massless case spin/(vector) 
direction can be either the direction of the 
electron motion  or the opposite. We then 
define for describing ``chirality''. It is 
usually distinguished by this quantity. That 
is to say $\gamma_5\Psi=+1 {\rm or} -1$.
Customary $+1$ is named Left moving- and $-1$ 
case is Right moving-Weyl or chiral fermion 
denoted $\Psi_L$ and $\Psi_R$ respectively. 
(The Lorentz or Poincare group of spacetime in $3+1$ dim Hermann Weyl investigated in detail 
and the basis is $2$ component spinor called 
Weyl spinors $\Psi_L$ and $\Psi_R$. In terms 
of 
these $4$ component Dirac field $\Psi$ such handed components can 
be constructed ($\Psi_D=\left(\begin{array}{c}\Psi_L\\\Psi_R\end{array}\right)$)

\section*{Appendix B}
We are now ready to discuss about 
Adler-Bell-Jacklin anomaly. In quantum 
field theory 
there are 
various symmetries. One of the most 
interesting symmetries is chiral (or axial) symmetry.
That is the interaction of Dirac field $\Psi_D$ with electromagnetic field $A_\mu$ is given by

\ 

$S=\int d^4x\bar{\Psi}(x)\left[i\gamma^\mu\left(\partial_\mu+ieA_\mu(x)\right)\right]\Psi_D(x) \ \ \ \ \ (\ast)$

\ \\
in the case of massless electron., where 
$\bar{\Psi}_D=\Psi^\dagger\gamma^0$.
It has chiral symmetry which may be 
obvious, 
if we rewrite ($\ast$) in terms of $\Psi_L$ and $\Psi_R$ as the Dirac eq. can be written as 

\ 

$\left(\begin{array}{cc}0&i(\partial_0+\sigma^i(\partial_i+ieA_i))\\i(\partial_0-\sigma^i(\partial_i+ieA_i))&0\end{array}\right)\left(\begin{array}{c}\Psi_L\\\Psi_R\end{array}\right)=0.$

 \ 

In this way the equations of $\Psi_L$ and $\Psi_R$ are separately given by the following Weyl equations

\ 

$i(\partial_0-\sigma^i(\partial_i+A_i))\Psi_L=0$\\
and

$i(\partial_0+\sigma^i(\partial_i+A_i))\Psi_R=0.$

\ 

In these forms it is evident that the 
theories are invariant under the following 
infinitesimal Weyl transformations

\ 

$\Psi_L\rightarrow (1-i\alpha^i\frac{\sigma_i}{2}-\beta^i\frac{\sigma_i}{2})\Psi_L$

$\Psi_R\rightarrow (1-i\alpha^i\frac{\sigma_i}{2}+\beta^i\frac{\sigma_i}{2})\Psi_R$

\ 

Where $\alpha^i$ and $\beta^i\ (i=1,2,3)$ are 
infinitesimal transformation parameters,
restricted to leave the normalization 
of the Weyl fields invariant. This Weyl or 
Chiral transformation is broken due to 
quantum effect in quantum field theory. There 
were several suggestive articles, but 
explicit manifestation is presented by
\begin{itemize}
\item S. Adler, Phys. Rev. 177 (1969) 2426 

and 

\item
 J. S. Bell and R. Jackiw Nuovo Cimento 60A (1969) 4.
\end{itemize}
Furthermore the method of path integral 
formulation this ABJ anomaly is due to nor-invariance of the path integral measure 
\begin{itemize}
\item K. Fujikawa Phys. Rev. Lett. 42 1195 (1979) 
\end{itemize}
Phenomenologically this ABJ anomaly is really 
important.
It has been observed by experiments. $\Pi^0$ 
meson decays into $2$ photons. When we 
approximate $\Pi^0$ as being massless, this 
decay process is expressed as the following 
diagram, triangle diagram of Feynman diagram
\begin{center}
\includegraphics[clip, width=10cm]{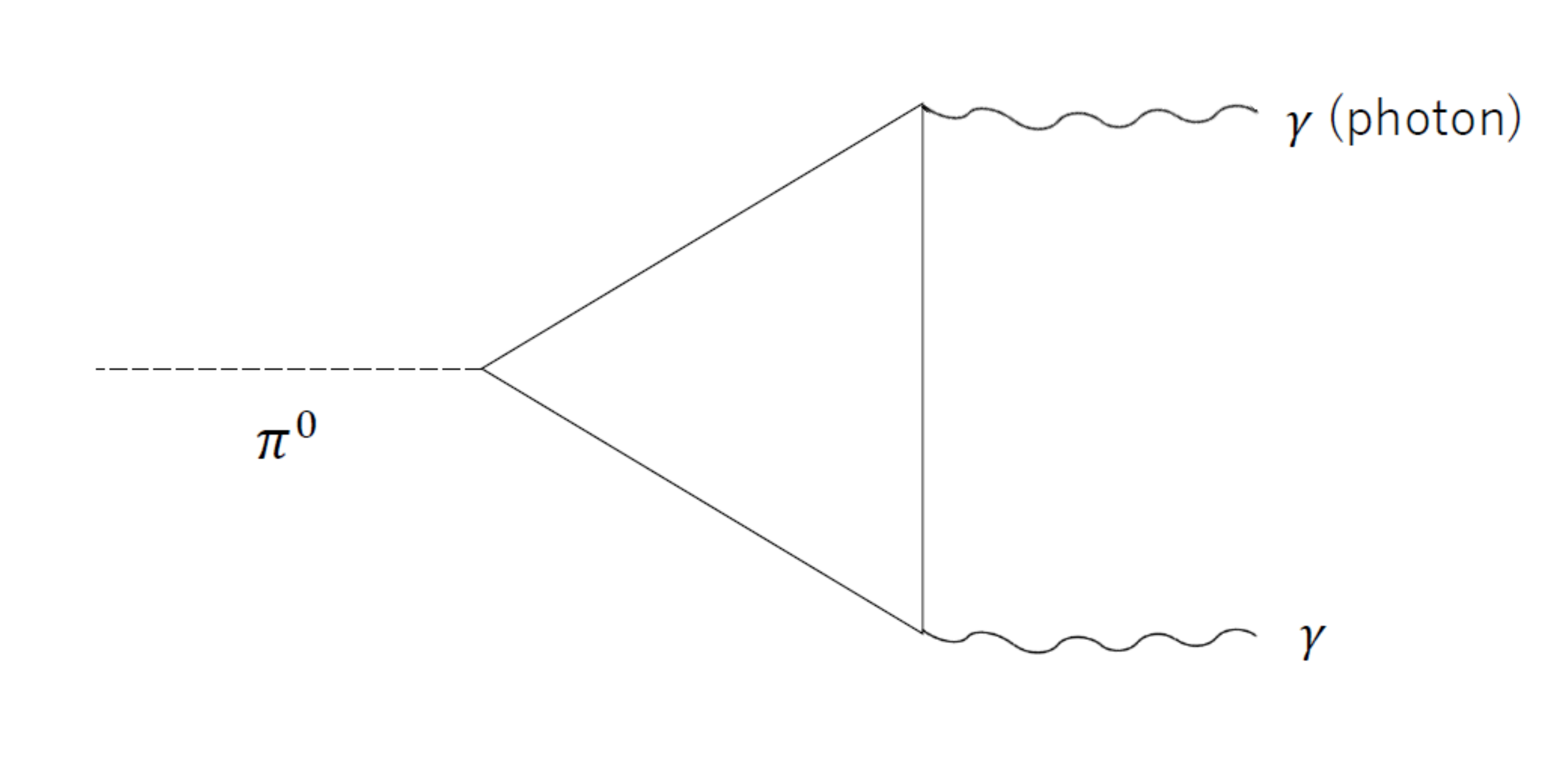}
\end{center}
If chiral symmetry is not broken, this 
diagram turns out to give zero. Thus this 
decay is not allowed. However, experimentally 
this decay process certainly exists. This is 
the evidence that Adler-Bell-Jackiw anomaly 
does exist.
In high energy, physics the ABJ anomaly is 
expressed as the non-conservation chiral 
current $J^5_\mu$ such that 

$\partial^\mu J^5_\mu=-\frac{e^2}{16{\pi}^2}\epsilon^{\alpha\beta\gamma\delta}F_{\alpha\beta}F_{\gamma\delta}$
\\
Here the chiral current $J^5_\mu$ is defined as

$J^5_\mu=\lim_{\epsilon\rightarrow 0}\left\{\bar{\Psi}(x+\frac{\epsilon}{2})\gamma_\mu \gamma^5{\rm exp}\left[ -ie\int^{x+\frac{\epsilon}{2}}_{x-\frac{\epsilon}{2}}dz A(z)\right]\Psi(x+\frac{\epsilon}{2})\right\}$

We might 
perform
the calculation to 
show that the above triangle diagram is 
non-zero 
due to the ABJ anomaly.
But we have instead in subsection
\ref{Paritynon} alluded to a derivation of the 
ABJ-anomaly by using how particles are pumped 
up or down from or to the fermi-sea (in high 
energy physics the Dirac sea).

\end{document}